\documentclass[10pt,journal]{IEEEtran}

\usepackage[english]{babel}
\usepackage[T1]{fontenc}
\usepackage[utf8]{inputenc}
\usepackage{xargs}

\usepackage{amssymb}
\usepackage{mathtools}
\usepackage{nccmath}

\usepackage{graphicx}
\usepackage[dvipsnames]{xcolor}

\ifCLASSOPTIONcompsoc
\usepackage[caption=false,font=normalsize,labelfont=sf,textfont=sf]{subfig}
\else
\usepackage[caption=false,font=footnotesize]{subfig}
\fi

\usepackage[export]{adjustbox}  

\usepackage[ruled,norelsize]{algorithm2e}
\makeatletter
\newcommand{\removelatexerror}{\let\@latex@error\@gobble}
\makeatother

\usepackage{etoolbox}
\usepackage{tikz}
\usetikzlibrary{tikzmark}
\usetikzlibrary{calc}

\usepackage{booktabs}
\usepackage{colortbl}
\usepackage{threeparttable}

\usepackage{siunitx}

\usepackage{hyperref}

\usepackage{xspace}
\usepackage{relsize}
\usepackage{bm}
\DeclarePairedDelimiter{\paren}{\lparen}{\rparen}
\DeclarePairedDelimiter{\parenEl}{[}{]}
\DeclarePairedDelimiter{\set}{\lbrace}{\rbrace}

\newcommand{\interval}[4]{\mathopen{#1}#2 \mathclose{}\mathpunct{}, #3\mathclose{#4}}
\newcommand{\intervalcc}[2]{\interval{[}{#1}{#2}{]}}

\DeclarePairedDelimiter{\inner}{\langle}{\rangle}

\newcommand\PS{{\mathrm{ps}}}

\newcommand\BG{\mathrm{bg}}
\newcommand\OBJECT{\mathrm{obj}}
\newcommand\CONVEX{\mathrm{co}}

\newcommand\MIN{\mathrm{min}}

\renewcommand{\vec}[1]{\bm{#1}}

\newcommand*\x{\ensuremath{\vec{x}}\xspace}

\newcommand*\y{\ensuremath{\vec{y}}\xspace}

\renewcommand*\r{\ensuremath{r}\xspace}

\newcommand*\dir{\ensuremath{\vec{d}}\xspace}

\newcommand\onesVec{\vec{1}}
\newcommand\zerosVec{\vec{0}}
\newcommand\indicatorVec{\vec{e}}

\renewcommand\matrix[1]{\bm{#1}}
\newcommand{\transpose}[1]{#1^T}
\newcommand{\inverse}[1]{\paren*{{#1}}^{-1}}
\DeclareMathOperator*{\diag}{diag}

\DeclarePairedDelimiter{\Lone}{\lVert}{\rVert_1}
\DeclarePairedDelimiterXPP\Ltwo[1]{}{\lVert}{\rVert}{_2}{#1}
\DeclarePairedDelimiterXPP\Linf[1]{}{\lVert}{\rVert}{_\infty}{#1}

\DeclarePairedDelimiter{\abs}{|}{|}

\newcommand*\R{\mathbb{R}}
\newcommand*\Rpos{\R_{+}}
\newcommand*\Z{\mathbb{Z}}
\newcommand*\Zpos{\Z_{+}}

\newcommand{\diff}{\mathop{\mathrm{{}d}}\mathopen{}}
\newcommand*\dx{\diff \x}
\newcommand*\dt{\diff \t}
\newcommand*\dy{\diff \y}

\newcommand*\dirac{\delta}
\newcommand\Poisson{\mathrm{Poisson}}

\DeclareMathOperator{\vectorise}{vec}
\newcommand\ConvSymbol{*}
\newcommand\conv{\mathbin{\ConvSymbol}}
\newcommand*{\defeq}{\mathrel{\coloneqq}}
\newcommand*{\eqdef}{\mathrel{\eqqcolon}}

\newcommand{\indicatorOptim}{\ensuremath{\iota}}

\newcommand\expe{\mathrm{e}}
\newcommand\psnr{\ensuremath{\mathrm{PSNR}}\xspace}
\newcommand\nll{\mathrm{nll}}
\newcommand\nllNormed[1][\PX]{\mathrm{nll}_{#1}}

\newcommand\bessOneZero{J_0}

\newcommand\supp{\operatorname{supp}}

\newcommand\params{{\vec{\theta}}}
\newcommand\paramsHat{\widehat{\params}}
\newcommand\paramSet{\vec{\Theta}}

\newcommand\px{\mathcal{P}}
\newcommand\iPx{\ensuremath{j}\xspace}
\newcommand\iPxSet{\mathcal{J}}
\newcommand\pxPos[1][\iPx]{\x_{#1}}

\newcommand\pxArea{\abs{\px}}

\renewcommand\t{\ensuremath{t}\xspace}

\newcommand\objMeas[1][\t]{\ensuremath{\phi_{#1}}\xspace}

\newcommand\intensSymbol{\varphi}
\newcommand\intens[1][\t]{\intensSymbol_{#1}}

\newcommand\bgPhi[1][\t]{\intens[#1]^{\BG}}

\newcommand\psPos[1][\t]{\x_{#1}^{\PS}}
\newcommand\psPosHat[1][\t]{\widehat{\x}_{#1}^{\PS}}
\newcommand\psPhi[1][\t]{\intens[#1]^{\PS}}
\newcommand\psPhiHat[1][\t]{\widehat{\intensSymbol}_{#1}^{\PS}}

\newcommandx\curveDir[2][1=, 2=\t]{\dir_{#2}^{#1}}

\newcommand\mapPxToImg{\nu}
\newcommand\mapImgToPx{\mapPxToImg^{-1}}

\newcommand\obsPhCount[1][\iFrame\iPx]{\ensuremath{N^\textrm{photon}_{#1}}\xspace}
\newcommand\obsPhCountVec[1][\iFrame]{\ensuremath{\vec{n}_{#1}}\xspace}
\newcommand\expPhCount[1][\iFrame\iPx]{\ensuremath{\mu_{#1}}\xspace}
\newcommand\expPhCountVec[1][\iFrame]{\ensuremath{\vec{\mu}_{#1}}\xspace}

\newcommand\obsGVCount[1][\iFrame\iPx]{\ensuremath{N^\textrm{grey}_{#1}}\xspace}
\newcommand\obsGVMat[1][\t]{\matrix{Z}_{#1}}

\newcommand\nSlices{{n^{\!\!\;\mathrm{s}}}}  
\newcommand\nHeight{{n^{\!\!\;\mathrm{h}}}}  
\newcommand\nWidth{{n^{\!\!\;\mathrm{w}}}}  
\newcommand\nPx{\ensuremath{{n^{\mathrm{p}}}}\xspace}

\newcommand\iFrame{k}

\newcommand\tA{\ensuremath{\t_a}\xspace}
\newcommand\tE{\ensuremath{\t_e}\xspace}
\newcommandx\tAin[1][1=t_{\iFrame}]{\ensuremath{\intervalcc{#1}{#1 + \tA}}\xspace}

\newcommand\psfFun{\kappa}
\newcommand\psfVec[1][\t]{\vec{\psfFun}_{#1}}

\newcommand\blurMat{\matrix{K}}

\newcommand\const{C}

\newcommand\BW{\mathrm{bw}}
\newcommand\NA{\mathrm{NA}}
\newcommand\refIndImm{n_i}
\newcommand\psfFunBW{\psfFun^{\BW}}
\newcommand\constBW{\const^{\BW}}

\newcommand\SG{\mathrm{sg}}
\newcommand\psfFunSG{\psfFun^{\SG}}
\newcommand\constSG{\const^{\SG}}
\newcommand\stdXY{\sigma_{xy}}
\newcommand\stdZ{\sigma_{z}}

\newcommand\GM{\mathrm{gm}}
\newcommand\psfFunGM{\psfFun^{\GM}}

\newcommand\dict{\mathcal{D}}
\newcommand\dictPosHat{\widehat{\mathcal{D}}^{\mathrm{thr}}}
\newcommand\iDict{d}

\newcommand\kerCovSet{\mathcal{C}}
\newcommand\nKerCov{{n^{\!\mathsmaller{\kerCovSet}}}}
\newcommand\iKerCov{k}
\newcommand\iKerCovTmp{\iKerCov'}
\newcommand\iKerCovSet{\mathcal{K}}
\newcommand\kerPosSet{\mathcal{X}}
\newcommand\nKerPos[1][\iKerCov]{n^\mathcal{X}_{#1}}
\newcommand\iKerPos{m}
\newcommand\iKerPosSet{\mathcal{M}}
\newcommand\wKerNormed[1][\iKerCov\iKerPos]{v_{#1}}
\newcommand\wKerNormedVec[1][\iKerCov]{\vec{v}_{#1}}
\newcommand\wKerNormedVecHat[1][\iKerCov]{\widehat{\vec{v}}_{#1}}
\newcommand\wKerNetVec[1][]{\vec{\intensSymbol}^{\mathrm{net}}_{#1}}
\newcommand\wKerNetVecHat[1][]{\widehat{\vec{\intensSymbol}}^{\mathrm{net}}_{#1}}

\newcommand\wKerNetVecDeb[1][]{\widehat{\vec{\intensSymbol}}^{\mathrm{deb}}_{#1}}
\newcommand\wKerMin{\intensSymbol_\MIN}
\newcommand\simplex{\Delta}

\newcommand\axialSampling{\Delta_{z}}
\newcommand\lateralSampling{\Delta_{xy}}
\newcommand\wavelength{\lambda}
\newcommand\quantEfficiency{q_{\wavelength}}
\newcommand\mGain{M}
\newcommand\ADU{f}
\newcommand\cameraBias{b}

\newcommand\imageDom{\Omega}

\DeclareMathOperator{\argminop}{\mathrm{arg\,min}}

\newcommand\argmin[1]{\underset{#1}{\argminop} \ }

\newcommand\st{\mathrm{s.t.}\ }

\newcommand\energy{\mathcal{E}}

\newcommand\eyeMat{\vec{I}}

\newcommand\regTerm{\mathrm{R}}
\newcommand\reg{\eta}
\newcommand\regVec{\vec{\eta}}
\newcommand\step{\gamma}

\newcommand\iIter{i}

\newcommand\operator{\vec{O}}

\DeclareMathOperator{\Prox}{Prox}
\DeclareMathOperator{\Proj}{Proj}

\renewcommand{\b}{\vec{b}}
\newcommand{\bc}{\vec{b}^i}
\newcommand{\bn}{\vec{b}^{i + 1}}
\newcommand\w{\vec{w}}
\newcommand{\wc}{\w^i}
\newcommand{\wn}{\w^{i + 1}}

\newcommand\wKer[1][]{\vec{\varphi}_{#1}}

\newcommand\wKern[1][]{\wKer[#1]^{\iIter + 1}}

\newcommand\rn{\vec{r}^{\iIter+1}}

\newcommand\wKerLS[1][\iKerCov]{\vec{\varphi}_{#1}}

\newcommand\wKerLSn[1][\iKerCov]{\wKerLS[#1]^{\iIter + 1}}

\makeatletter
\newcommand\ensuresingleperiod{\@ifnextchar.{}{.\@\xspace}}
\makeatother

\newcommand\ie{\textit{i.e.}\xspace}
\newcommand\eg{\textit{e.g.}\xspace} 

\newcommand\viceversa{\textit{vice versa}\xspace}
\newcommand\versus{\textit{vs.}\xspace}

\newcommand\aposteriori{\textit{a posteriori}\xspace}

\renewcommand\refeq[1]{\eqref{#1}}
\newcommand\reffig[1]{Fig.~\ref{#1}}
\newcommand\refsubfig[1]{\protect\subref{#1}}
\newcommand\refalg[1]{Algorithm~\ref{#1}}
\newcommand\reftable[1]{Table~\ref{#1}}
\newcommand\refsection[1]{Section~\ref{#1}}

\usepackage{amsthm}

\newtheorem*{result*}{Result}
\usepackage{layouts}

\title{Modelling point spread function in fluorescence microscopy with a sparse
Gaussian mixture: trade-off between accuracy and efficiency}

\author{Denis~K.~Samuylov, Prateek~Purwar, Gábor~Székely, and~Grégory~Paul}

\begin{document}
\maketitle
\begin{abstract}
%
Deblurring is a fundamental inverse problem in bioimaging.
It requires modelling the point spread function (PSF), which captures the
optical distortions entailed by the image formation process.
The PSF limits the spatial resolution attainable for a given microscope.
However, recent applications require a higher resolution, and have prompted the
development of super-resolution techniques to achieve sub-pixel accuracy.
This requirement restricts the class of suitable PSF models to analog ones.
In addition, deblurring is computationally intensive, hence further requiring computationally efficient models.
A custom candidate fitting both requirements is the Gaussian model.
However, this model cannot capture the rich tail structures found in both
theoretical and empirical PSFs.
In this paper, we aim at improving the reconstruction accuracy beyond the
Gaussian model, while preserving its computational efficiency.
We introduce a new class of analog PSF models based on Gaussian mixtures.
The number of Gaussian kernels controls both the \emph{modelling accuracy} and
the \emph{computational efficiency} of the model: the lower the number of
kernels, the lower accuracy and the higher efficiency.
To explore the accuracy--efficiency trade-off, we propose a variational
formulation of the PSF calibration problem, where a convex sparsity-inducing
penalty on the number of Gaussian kernels allows trading accuracy for
efficiency.
We derive an efficient algorithm based on a fully-split formulation of
alternating split Bregman.
We assess our framework on synthetic and real data and demonstrate a better
reconstruction accuracy in both geometry and photometry in point source
localisation---a fundamental inverse problem in fluorescence microscopy.
%
%

\end{abstract}




\section{Introduction}
\label{sec:psf:intro}
\IEEEPARstart{I}{ncoherent} imaging systems are linear in power (\ie intensity)
and are well described by linear system theory: the mean intensity image results
from the \emph{superposition integral} of the true object intensity weighted by
the \emph{point spread function}~(PSF).
The PSF encodes the propagation of the incoherent light emitted by a
unit-intensity point source object through the optics up to the array of
photo-detectors.
In turn, the mean intensity parameterises the statistical model that accounts
for the intrinsic stochastic nature of light emission (see \eg \cite{Snyder1993,
Chesnaud1999, Paul2013}).
Together with other sources of distortion, this defines the image formation
process, \ie the \emph{forward problem}.

Solving the \emph{inverse problem} for an incoherent imaging system amounts to
reconstructing the number (amount), the position (geometry), and the intensity
(photometry) of the emitting light sources.
Depending on the nature of the reconstruction space, \emph{digital} or
\emph{analog}, the task is qualitatively different.
In the digital case, the number and position of the light sources in physical
space are lost: only the integrated intensity at each pixel position is
reconstructed.
On the contrary, an analog reconstruction aims at resolving the number, the
position, and the intensity of the emitting light sources \emph{in physical
  space}.
The PSF and the inherent noise in the acquisition process make this inverse
problem ill-posed and ill-conditioned.
Therefore, ``good'' models about the forward problem and the imaged objects are
necessary for solving this problem.
In this work, we focus on modelling the first step of the image formation
process by finding a PSF representation, suitable for solving imaging inverse
problems in an \emph{analog reconstruction space}.

%
%
The quality of the PSF model determines the accuracy of the imaged object
reconstruction, thus its knowledge is fundamental for solving bio-imaging
inverse problems \cite{Scalettar1996, Mcnally1999, Haeberle2001}.
One approach is to assume that the PSF is unknown and to include its estimation
in the inverse problem (blind deconvolution, see \eg
\cite{Pankajakshan2009BlindDeconv, Hadj2013}).
However, it makes the problem more ill-conditioned.
%
%
Therefore, we focus on estimating the PSF as a separate task.
We also restrict ourselves to shift-invariant imaging systems, excluding
space-varying PSF models.

PSF models can be classified according to different criteria: \emph{model-based}
\versus \emph{phenomenological}, \emph{digital} \versus \emph{analog}, and
\emph{calibrated} \versus \emph{uncalibrated}.
The PSF model can be derived either from modelling the imaging systems or from
imaging point-like objects.
A digital PSF is well-suited for a digital reconstruction space, but not adapted
to analog reconstructions; whereas an analog PSF can be used both in analog and
digital reconstructions (after appropriate integration and sampling).
Finally, if the PSF model is calibrated experimentally on a specific imaging
system, we call this PSF \emph{calibrated}.
These criteria define a range of possible models.
At the two extremes stand the theoretical (derived from optics) and the
empirical (punctual source measurements) PSF models. The former is analog and
can potentially achieve sub-pixel resolution, but at the expense of costly
convolution algorithms.
On the contrary, the latter is digital and therefore amenable to efficient
convolution algorithms (based on the fast Fourier transform, FFT), but at the
expense of a loss in accuracy.

Theoretical PSF models are derived from modelling the microscope optics.
Their differences lie in their level of simplification.
The simplest models use the paraxial approximation, which is valid for
objectives with a small numerical aperture (NA), thin imaged specimens, and
ideal acquisition conditions \cite{Born2000}.
%
%
More advanced models account for the spherical aberrations due to sample
thickness and refractive indexes mismatch \cite{Gibson1992}.
%
Recent models use the vectorial theory of diffraction to extend PSF models to
objectives with a high NA \cite{Hell1993, Torok1997, Haeberle2002, Ghosh2015}.
In practice, this class of PSF models is difficult to use because hard or
impossible to calibrate (\eg unknown refractive index of the immersion medium
\cite{Haeberle2001}).
In addition, a specific imaging system can deviate from these idealised models
in at least two ways (\eg \cite{Gibson1992, Sibarita2005}):
the experimental setup does not fit the model assumptions or variations in the
optics (\eg lens imperfections, optics misalignment) create distortions not
accounted for by the models.

Empirical PSF models are derived from \emph{calibration data}, namely an image
stack of point-like objects, such as quantum dots, fluorescent beads (\eg
\cite{Agard1989, Hiraoka1990, Shaw1991, Monvel2001, Juskaitis2006})
or small structures within the sample \cite{Monvel2003}.
Calibration data provide an accurate representation of the PSF only for specific
imaging conditions \cite{Agard1989}.
Any deviation from the expected PSF image can be used to identify problems in
the experimental setup (\eg oil mismatch, vibrations from the heating system
\cite{Sibarita2005, Juskaitis2006}).
%
However, it also imposes a restriction: calibration data must be acquired under
the same conditions as the specimen, \ie from point-like objects embedded within
the sample \cite{Mcnally1999,Sibarita2005}.
This is rarely feasible in practice
and may require advanced measurement techniques \cite{Shaevitz2007}.
Moreover, empirical PSF models require correcting the corruptions entailed by
the image formation process, \eg by removing the background signal, denoising,
or accounting for the extended geometry of point-like objects (\eg
\cite{Mcnally1994, Scalettar1996, Sibarita2005, Pankajakshan2009}).

The two ends of the spectrum of possible PSF models illustrate a trade-off
between accuracy and efficiency.
On one end, theoretical PSF models can be used to resolve the localisation of
individual point sources in physical space, at the expense of robustness
(difficulty to fit a specific setup) and computational efficiency (analog PSF
are typically more expensive to evaluate than digital ones).
On the other end, digital PSF models are computationally efficient to use (\eg
thanks to fast algorithms such as FFT), but their resolution is limited by the
pixel grid.
Between these two extremes, one can trade accuracy for efficiency or \viceversa.
For example, theoretical PSF models can be adjusted to a specific setup by using
PSF measurements (\eg \cite{Haeberle2001, Hanser2004, Pankajakshan2009}).
%
%
This improves robustness, but does not solve the computational efficiency
problem.
For this purpose, a well-spread solution is to use a Gaussian PSF model
approximating the real PSF of a specific imaging system (\eg \cite{Zhang2007}).
Such models are well suited for modelling the in-focus section of the PSF
\cite{Aguet2009}, but results in a poor approximation of the PSF tails
\cite{Small2014} and are not suitable for modelling the 3D PSF of a widefield
microscope \cite{Zhang2007}.
In order to explore the trade-off between accuracy and efficiency, a solution is
to use \emph{analog phenomenological models}.
These models are flexible enough to accommodate real PSF shapes, and still
computationally efficient. Moreover, their complexity can be adjusted according
to the desired trade-off between accuracy and efficiency.
For example, Zernike moments can encode reflection-symmetric PSFs
\cite{Bissantz2010, Maalouf2011} with a complexity controlled by the number of
moments used in the model.
More recently, polynomial B-splines were proposed \cite{Kirshner2013Splines,
Tahmasbi2015}.
In that case, the trade-off can be explored by varying the number and placement
of the basis functions.
Nonetheless, to the best of our knowledge, this trade-off has not yet been
systematically explored in the literature.

In this paper, we aim at improving the PSF reconstruction accuracy beyond single
Gaussian models that misapproximates the PSF tails \cite{Small2014}.
Therefore, we propose a new class of phenomenological PSF models with an
adjustable level of complexity that can be tuned to achieve a desired
accuracy--efficiency trade-off.
We model the PSF as a sparse mixture of multivariate Gaussian distributions.
The motivation behind this choice is a fast algorithm to compute non-gridded
convolutions with Gaussian kernels: the improved fast Gaussian transform (IFGT,
\cite{Yang2003}).
%




\section{Forward problem}
\label{sec:psf:fwd}


\begin{figure*}[!t]
  \centering
  \includegraphics[width=\textwidth]{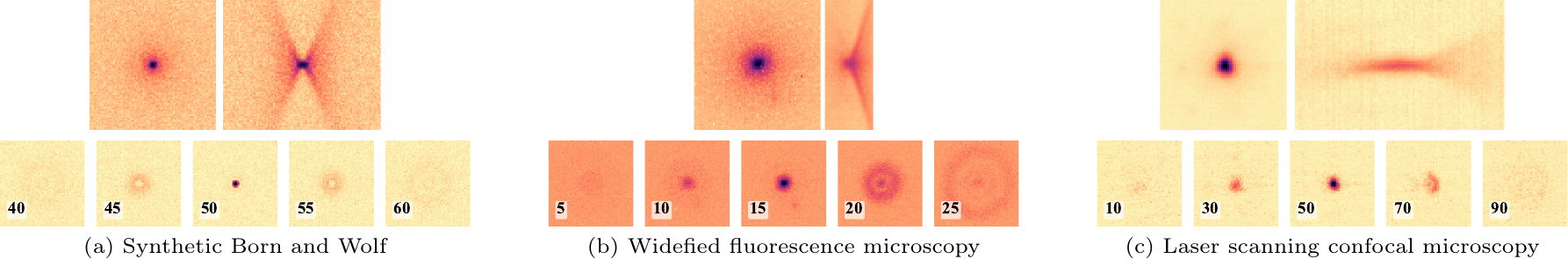}
  \subfloat{\label{fig:psf:psf-gallery:SBW}}
  \subfloat{\label{fig:psf:psf-gallery:WFFM}}
  \subfloat{\label{fig:psf:psf-gallery:LSCM}}
  \caption{
    \textbf{Fluorescent bead image data for different imaging modalities.}
    \refsubfig{fig:psf:psf-gallery:SBW} Synthetic image stack of an idealised PS
    object using the forward model in \refsection{sec:psf:fwd} and the BW PSF
    model.
    \refsubfig{fig:psf:psf-gallery:WFFM}-\refsubfig{fig:psf:psf-gallery:LSCM}
    Real image data of nanoscale beads acquired using a \emph{widefied
    fluorescence microscopy} \refsubfig{fig:psf:psf-gallery:WFFM} and a
    \emph{laser scanning confocal microscopy}
    \refsubfig{fig:psf:psf-gallery:LSCM}.
    Orthogonal z- and x- mean projections (\emph{first row}) and sample focal
    planes (\textit{second row}) shown in $\log$-colour scale.}
  \label{fig:psf:psf-gallery}
\end{figure*}


To ease the reading of the presentation of the forward problem we provide in
Table~\ref{table:psf:notations} a summary of the notations used
in the forthcoming sections in the appendix.

\subsection{Object model}
\label{sec:psf:point-source-object}

We assume that the PSF measurements result from imaging a fluorescent bead with
a size below the resolution limit of the microscope objective, thus
approximating an idealised \emph{point source}~(PS) object~\cite{Sibarita2005}.
We further assume that during acquisition, the fluorescent bead is immobile and
emits photons at a constant rate (\ie we neglect photo-bleaching).
The total photon flux decomposes as the contribution of the photons emitted by
the PS located at $\psPos[] \in \R^3$ at rate $\psPhi[] \in \Rpos$ and the
photons emitted by the background (due to autofluorescence) at a constant rate
$\bgPhi[] \in \Rpos$.
We use the measure-theoretical framework of \cite{Samuylov2015} to model the
total photon flux as a spatio-temporal object measure:
\begin{equation}
  \label{eq:psf:ps-obj}
  \objMeas[]^\OBJECT\paren{\dy \times \dt} = \paren*{\psPhi[] \dirac_{\psPos[]}\paren{\dy} +\bgPhi[]\dy} \dt \ .
\end{equation}

\subsection{Image formation process}
\label{sec:psf:image-formation-model}
The image formation in fluorescence microscopy can be modelled as a two-stage
process~\cite{Samuylov2015,Snyder1993}.
The \emph{object-to-pixel mapping} models the optical distortions due to the
random nature of light emission and propagation: it integrates the expected
photon flux emitted by a set of fluorescent objects located in physical space
over the pixel surface during the exposure time and models the number of
photons hitting each photo-detector.
The \emph{pixel-to-image mapping} models the conversion of photon counts
collected at each pixel to image grey values and accounts for the measurement
noise, \eg due to the photoelectric effect, signal amplification, noise in the
electronic circuitry and quantisation \cite{Hirsch2013,Konnik2014}.

\subsubsection{Object-to-pixel mapping}
The PSF tails are in the low-intensity region where the fluctuations inherent to
light emission are measurable and photon shot-noise is significant.
Therefore, we use a Poissonian model for the photon counting statistics.
We denote the imaging volume as $\imageDom \subset \R^3$ and model the photon
count at pixel $\px_\iPx \subset \imageDom$ during the exposure time $\tE$ as a
random variable $\obsPhCount[\iPx]$ following a Poisson distribution with mean
intensity $\expPhCount[\iPx]$:
\begin{equation}
  \label{eq:psf:noise-model}
  \obsPhCount[\iPx] \sim
  \Poisson\paren{\expPhCount[\iPx]} \ .
\end{equation}

In incoherent imaging, the mean intensity at a given location in the image plane
results from the superposition of the independent contributions of the light
sources distributed in physical space.
For each individual light source, the photon flux observed under a specific
imaging system is entirely characterised by the PSF.
After a proper normalisation, the PSF can be understood microscopically as the
transition probability kernel, denoted $\psfFun\paren{\dx \mid \y}$, translating
individual photons in the image plane around $\dx$, conditionally on the
position $\y$ of the emitting light
source~\cite{Richardson1972,Snyder1993,Thiebaut2016}.
%

For a general object modelled as a spatio-temporal measure, denoted
$\objMeas[]$, the expected photon count at $\iPx$-th pixel is described as a
superposition integral:
\begin{equation}
  \label{eq:psf:superposition-flux}
  \expPhCount[\iPx]\paren{\objMeas[]} =
  \int_{\px_\iPx \times \intervalcc{0}{\tE}}\int_\imageDom \psfFun\paren{\dx \mid \y} \, \objMeas[]\paren{\dy \times \dt}\ .
\end{equation}
We assume that the imaging setup is well described by a shift-invariant PSF. In
this case, the superposition integral becomes a convolution integral:
\begin{equation}
  \label{eq:psf:superposition-convolution}
  \int_\imageDom \psfFun\paren{\dx \mid \y} \, \objMeas[]\paren{\dy \times \dt} =
 \paren*{\psfFun\conv\objMeas[]}\paren{\dx \times \dt}\ .
\end{equation}
For a set of PSs with position and intensity $\set{\paren{\x_s, \intens[s]}}_s$,
equation~\refeq{eq:psf:superposition-convolution} transforms in a linear combination
of shifted PSF:
\begin{equation*}
  \paren*{\psfFun \conv \sum_s \intens[s] \dirac_{\x_s}}(\dx \times \dt) =
  \paren*{\sum_s \intens[s] \psfFun(\x-\x_s)}\dx\dt\ .
\end{equation*}
We specialise the above equation to the object model~\refeq{eq:psf:ps-obj} and
use a mid-point quadrature to approximate the photon
flux~\refeq{eq:psf:superposition-flux}:
\begin{equation}
  \label{eq:psf:img-exp}
  \expPhCount[\iPx]\paren{\objMeas[]} \approx
  \expPhCount[\iPx]\paren{\psPos[], \psPhi[]} \defeq
  c\, \paren*{\psPhi[]\psfFun \paren{\x_\iPx - \psPos[]} + \bgPhi[]}\ ,
\end{equation}
where $\x_{\iPx} \in \imageDom$ is the centre of the $\iPx$-th pixel, $c \defeq
\pxArea \tE$ the spatio-temporal integration volume, and $\pxArea$ the pixel
area.
For clarity we have omitted the dependence on $\bgPhi[]$, that is estimated
independently in practice.
From now on, we rescale the photon emission rates by $c$ to
obtain the integrated intensities, that we denote for simplicity with the same
symbols: $\psPhi[] \defeq c\, \psPhi[]$ and $\bgPhi[] \defeq c\, \bgPhi[]$.
We assume that the pixels are back-projected onto the object space, \ie the
lateral pixel coordinates are divided by the magnification factor and the axial
coordinate is aligned with the object \cite{Aguet2009}.

\subsubsection{Pixel-to-image mapping}
\label{sec:psf:map:pixel-image}
Modelling the conversion of photons hitting the detector surface into grey
values depends on the type of camera \cite{Hirsch2013,Konnik2014}.
We model the conversion by a deterministic affine mapping, denoted
$\mapPxToImg$:
\begin{equation}
  \label{eq:psf:px2img}
  \obsGVCount[\iPx]  =
  \quantEfficiency\, \mGain\, \ADU^{-1} \obsPhCount[\iPx]\, + \cameraBias \eqdef
  \mapPxToImg\paren{\obsPhCount[\iPx]}\ ,
\end{equation}
where $\quantEfficiency$ is the quantum efficiency at emission wavelength
$\wavelength$, $\mGain$ is the multiplication gain, $\ADU$ is the
analog-to-digital proportionality factor, and $\cameraBias$ is the camera bias.

\subsection{PSF models}

We introduce three PSF models spanning the complexity range.
The classical Born and Wolf~(BW) model is used as a reference idealised model,
assuming that the microscope is operated under design conditions.
The single Gaussian~(SG) model is a widespread PSF approximation because of its
computational advantages.
The Gaussian mixture~(GM) model, which we propose in this work, represents a
whole class of models that will allow us exploring the accuracy--efficiency
trade-off.
Each model is parameterised by a set of parameters, generically denoted
$\params$, belonging to a set of admissible values, denoted $\paramSet$. When
ambiguous, we will add a superscript to $\params$ and $\paramSet$ with the PSF
model name.
We denote by $\x \defeq (x,y,z) \in \R^3$ an arbitrary point in physical space.

\subsubsection{Born and Wolf (BW) model}
The model is parameterised by three parameters, \ie the emission wavelength, the
numerical aperture, and the refractive index of the immersion medium, denoted
$\params^\BW \defeq \set{\wavelength, \NA, \refIndImm} \in \paramSet^\BW \defeq
\Rpos^3$. The BW PSF kernel writes as:
\begin{equation*}
  \label{eq:psf:model-bw}
  \psfFunBW_\params\paren{\x} \defeq \constBW_\params \Big| \int_0^1 \bessOneZero
  \paren{k_0\,\r\,\NA\,\rho } \,
  \expe^{-i \, \Phi\paren{\rho, z}} \rho \diff
  \rho\Big|^2 \ ,
\end{equation*}
where $\constBW_\params$ is the normalisation constant, $\bessOneZero$ denotes
the Bessel function of the first kind of order zero, $\Phi\paren{\rho, z} =
\frac{k_0 z \NA^2 \rho^2}{2 \refIndImm}$ is the phase term, $k_0 =
\frac{2\pi}{\wavelength\refIndImm}$ is the wavenumber in vacuum, and $\r =
\sqrt{x^2 + y^2}$.

\subsubsection{Single Gaussian (SG) model}
The SG model is simply a Gaussian density, with a particular covariance matrix:
this model is radially symmetric around the $z$-axis. Therefore, it is
parameterised by the lateral and axial standard deviations, denoted $\params^\SG
\defeq (\stdXY, \stdZ) \in \paramSet^\SG \subset \Rpos^2$. The SG PSF kernel is
defined as $\psfFunSG_{\params}\paren{\x} = \constSG_\params \,
\expe^{-\frac{1}{2}\x^T\Sigma_{\params}^{-1}\x}$,
%
%
where $\Sigma_{\params} \defeq \diag\paren*{\stdXY^2, \stdXY^2, \stdZ^2} \in
\R^{3\times3}$ is the covariance matrix. The normalisation constant is given by
$\constSG_\params = \paren*{8\pi^3\stdXY^4\stdZ^2}^{-0.5}$.

\subsubsection{Gaussian mixture (GM) model}
\label{sec:psf:model:gm}
%
\paragraph{Definition}
To explore the accuracy--efficiency trade-off, we introduce a Gaussian mixture
model, namely a convex combination of Gaussian kernels shifted to different
positions.
The general GM model is described by a \emph{parametric dictionary}, denoted
$\dict$, where each atom is a shifted SG kernel: $\dict \defeq
\bigcup_{(\params_a, \vec{x}_a) \in \params^\dict}
\psfFunSG_{\params_a}(\cdot-\x_a)$.
%
%
However, in this paper we structure $\params^\dict$ by a set of $\nKerCov$
covariance matrices:
for each $\params^\SG_\iKerCov$ with $\iKerCovSet \defeq \set{1, \dots,
\nKerCov}$ we associate a set of positions, defined as $\kerPosSet_\iKerCov
\defeq \bigcup\limits_{m \in \iKerPosSet_\iKerCov} \x_{\iKerCov\iKerPos}$ with
$\iKerPosSet_\iKerCov \defeq \set{1, \dots, \nKerPos}$: $\params^\dict \defeq
\bigcup_{\iKerCov \in \iKerCovSet} \set{\params^\SG_\iKerCov} \times
\kerPosSet_\iKerCov \eqdef \bigcup_{\iKerCov \in \iKerCovSet}
\params^\dict_\iKerCov$.
%
%
The dictionary size corresponds to the number of Gaussian kernels in the
mixture, defined as $\abs{\dict} = \sum_{\iKerCov\in\iKerCovSet} \nKerPos$.
To ensure the proper normalization of the PSF, a GM model is defined as a
\emph{convex combination} of the Gaussian kernels in the dictionary $\dict$.
The mixture weight vector is denoted $\wKerNormedVec[]$ and belongs to the
$\abs{\dict}$-simplex, denoted $\simplex_{\abs{\dict}}$, \ie $\wKerNormedVec[]$
must have positive components summing to one.
The parameters are the dictionary parameters and the mixture weights vector, \ie
$\params^\GM \defeq \params^\dict \cup \set{\wKerNormedVec[]}$.
The GM model then writes:
\begin{equation}
  \label{eq:psf:model-mg}
  \psfFunGM_\params\paren{\x} = \sum_{\iKerCov \in \iKerCovSet}
  \sum_{\iKerPos \in \iKerPosSet_\iKerCov} \wKerNormed\,
  \psfFunSG_{\params_\iKerCov} \paren{\x - \x_{\iKerCov\iKerPos}} \ .
\end{equation}
To capture the structure of the dictionary parameter set $\params^\dict$, we
regroup the kernels in $\dict$ and the mixture weights into
$\nKerPos$-dimensional vectors, denoted $\psfVec[\iKerCov]\paren{\x}$ and
$\wKerNormedVec$ respectively. Their \iKerPos-th elements are the translated
kernel
$\parenEl{\psfVec[\iKerCov]\paren{\x}}_\iKerPos \defeq
\psfFunSG_{\params_\iKerCov} \paren{\x - \x_{\iKerCov\iKerPos}}$ and the mixture
weight $\parenEl{\wKerNormedVec}_\iKerPos \defeq \wKerNormed$ respectively.
Then, the GM model~\refeq{eq:psf:model-mg} writes:
\begin{equation}
  \label{eq:psf:model-mg-vec} \psfFunGM_{\params}\paren{\x} =
  \sum_{\iKerCov\in\iKerCovSet} \transpose{\psfVec[\iKerCov]\paren{\x}}
  \wKerNormedVec \ .
\end{equation}
We define
$\supp(\wKerNormedVec[]) \defeq \set{(\iKerCov,\iKerPos) \in
  \iKerCovSet\iKerPosSet \mid \parenEl{\wKerNormedVec[]}_{\iKerCov\iKerPos} >
  0}$, the \emph{support of the mixture weights}, and
$\iKerCovSet\iKerPosSet \defeq \bigcup_{\iKerCov\in\iKerCovSet} \set{k}\times
\iKerPosSet_\iKerCov$.
For a fixed dictionary size, the support of $\wKerNormedVec[]$ sets the
\emph{effective computational effort} to evaluate the GM model by reducing the
number of SG kernel evaluations (\ie the sum in equation~\refeq{eq:psf:model-mg}
runs only over $\supp(\wKerNormedVec[])$).
The size of the support defines our \emph{measure of efficiency} for the MG
model.
If the set of kernel positions
$\kerPosSet \defeq \cup_{\iKerCov\in\iKerCovSet} \kerPosSet_\iKerCov$ is aligned
with the pixel centres, we call the dictionary \emph{digital}, otherwise we call
it \emph{analog}.

\paragraph{Virtual source interpretation}
We can interpret the GM model within the virtual microscope (VM) framework
\cite{Samuylov2015}.
The MG model~\refeq{eq:psf:model-mg} can be rewritten as a superposition of
$\nKerCov$ convolutions:
\begin{equation*}
  \psfFunGM_\params(\x)\dx\dt = \sum_{\iKerCov\in\iKerCovSet}
  \paren*{\psfFunSG_{\params_\iKerCov} \conv \sum_{\iKerPos\in\iKerPosSet_\iKerCov} \wKerNormed[\iKerCov\iKerPos]\dirac_{\x_{\iKerCov\iKerPos}}}(\dx\times \dt) \ .
\end{equation*}
Using the terminology of \cite{Samuylov2015}, the $\nKerPos[\iKerCov]$ light
sources modelling the $\iKerCov\text{-th}$ object measure (located at
$\x_{\iKerCov\iKerPos}$ and of intensity $\wKerNormed[\iKerCov\iKerPos]$) are
\emph{virtual}, in the sense that they do not represent physical light sources,
but they approximate a photon flux.
This observation leads to a representation of the convolution of the GM model
with an arbitrary object that is well-suited for the VM
framework~\cite{Samuylov2015}:
the GM model amounts to approximating the PSF by $\nKerCov$ optical systems
modelled using $\psfFunSG_{\iKerCov}$.
For each system, each point source $(\psPos[], \psPhi[])$ is replaced by
$\nKerPos$ point sources: $\set*{(\psPos[]+\x_{\iKerCov 1}, \wKerNormed[\iKerCov
1]\,\psPhi[]), \dots, (\psPos[]+\x_{\iKerCov \nKerPos}, \wKerNormed[\iKerCov
\nKerPos]\,\psPhi[])}$ that are convolved independently with
$\psfFunSG_{\iKerCov}$.
The resulting mean images are summed to produce the mean image of the original
imaging system.
%
%

\paragraph{Computational efficiency}
The GM model can be efficiently evaluated by the FFT or the IFGT
(see~\cite{Morariu2008}).
The FFT is suitable for the digital scenario when the object model and the PSF
model are sampled on the same grid.
The IFGT allows fast evaluation of the forward model based on the virtual source
formulation discussed above.
In both scenarios, the computational efficiency of the evaluation of the
Gaussian mixture PSF model depends on the size of its support.
%

\section{Inverse problem}
\label{sec:psf:inv}
We aim to estimate the parameters of the PSF models from calibration data.
In scanning fluorescence microscopy, the image stack is acquired by moving the
focal plane over the imaging volume \cite{Sibarita2005}.
The camera detector of size $\nHeight \times \nWidth$ pixels measures the
incident light intensity during the exposure time $\tE$ and encodes it into grey
values at $\nSlices$ positions of the focal plane.
The pixel size and the displacement between two consecutive focal planes are
denoted $\lateralSampling$ and $\axialSampling$, respectively.
As a result, the image stack is an array of grey values, denoted
$\obsGVMat[] \in \Zpos^{\nSlices\times\nHeight\times\nWidth}$, holding
measurements from $\nPx \defeq \nSlices\nHeight\nWidth$ pixels.
We make the standard assumption that the images in a stack are acquired
simultaneously in time.
%

\subsection{Estimation problem: maximum \aposteriori formulation}
\label{sec:psf:estimation}

\subsubsection{Problem formulation}
\label{sec:psf:MAP}
Given a PSF model $\psfFun_{\params}$, we solve a maximum \emph{a posteriori}
(MAP) optimisation problem to estimate $\params$.
In general, the position $\psPos[]$ and intensity $\psPhi[]$ of the PS object
are also unknown.
Therefore, the general inverse problem aims at jointly estimating the parameters
of the PS and the PSF models:
\begin{equation}
  \label{eq:psf:optim-general}
  \begin{aligned}
    \argmin{\psPos[],\, \psPhi[],\, \params} & \nll\paren{
      \psPos[], \psPhi[], \params \mid \obsGVMat[]}
    + \reg \, \regTerm\paren{\psPos[], \psPhi[], \params} \\
    \st & \psPos[] \in \imageDom, \ \psPhi[] \in \Rpos,\ \params
    \in \paramSet
  \end{aligned}
\end{equation}
where $\nll$ is called the data-fitting term, $\regTerm$ the regularisation term
encoding our prior knowledge about the model parameters, and $\reg$ the
regularisation parameter controlling the trade-off between them.
We call the problem~\refeq{eq:psf:optim-general} the \emph{parametric blind PS
deconvolution}. It specialises to two problems:

\paragraph{PS localisation}
The PSF is assumed known, and only the position and intensity of the PS are
estimated. In this case, the data-fitting term reduces to
$\nll\paren{\psPos[], \psPhi[] \mid \obsGVMat[], \params}$, and the PSF model
parameters are fixed.

\paragraph{PSF calibration}
The PS is assumed known, and only the PSF model parameters are
estimated. The data-fitting term reduces to
$\nll\paren{\params \mid \obsGVMat[], \psPos[], \psPhi[]}$, and the PS intensity
and position are fixed.

\subsubsection{Data-fitting functional}
In a Bayesian setting, the data-fitting term corresponds to the negative
log-likelihood~($\nll$) derived from the stochastic image formation model
described by~\refeq{eq:psf:noise-model}.
We convert the grey values into what we call the \emph{raw photon counts} using
\refeq{eq:psf:px2img} and vectorise it into
$\obsPhCountVec[] \defeq \vectorise\paren*{\mapImgToPx\paren{\obsGVMat[]}} \in
\Rpos^{\nPx}$.
Specialising the object-to-pixel mapping defined by \refeq{eq:psf:img-exp} for the
PSF model parameterised by $\params$, we compute the \emph{expected photon
count} at each pixel $\iPx \in \set{1, \dots, \nPx} \eqdef \iPxSet$ and stack
the results into a vector, denoted $\expPhCountVec[]\paren{\psPos[], \psPhi[],
\params} \in \Rpos^{\nPx}$.
To compare different models, we use the normalised version of the $\nll$, also
called \emph{deviance} or \emph{Bregman divergence} \cite{Paul2013}, which is
zero when over-fitting (\ie the model prediction at each pixel is its raw photon
count):
\begin{equation*}
  \begin{multlined}
    \nll(\psPos[], \psPhi[], \params \mid \obsGVMat[]) \eqdef \nll(\psPos[], \psPhi[], \params \mid \obsPhCountVec[]) \\
    = \Big< \onesVec_{\nPx}, \ \obsPhCountVec[]
    \log{\frac{\obsPhCountVec[]}{\expPhCountVec[]\paren{\psPos[],
          \psPhi[], \params}}} + \expPhCountVec[]\paren{\psPos[],
      \psPhi[], \params} - \obsPhCountVec[] \Big>\ .
  \end{multlined}
\end{equation*}

\paragraph{Vector of expected photon counts}
\label{sec:psf:vec-exp-pc}
The vector of expected photon counts is computed using~\refeq{eq:psf:img-exp}.
For the BW and SG models, the formulation is straightforward. However, the GM
model requires more attention in its formulation.
Substituting~\refeq{eq:psf:model-mg-vec} into~\refeq{eq:psf:img-exp}, we
evaluate the expected photon count at each pixel $\iPx \in \iPxSet$ and stack
the results into a vector:
\begin{equation}
  \label{eq:psf:ifm-gm-vec}
  \expPhCountVec[] \paren{\psPos[], \psPhi[], \params^\GM}
  = \psPhi[] \sum_{\iKerCov\in\iKerCovSet} \blurMat_\iKerCov\paren{\psPos[]} \, \wKerNormedVec + \bgPhi[]\,\onesVec_\nPx \ ,
\end{equation}
where $\blurMat_\iKerCov$ is the blurring matrix defined for the $\iKerCov$-th
kernel~as:
\begin{equation*}
  \transpose{\blurMat_\iKerCov\paren{\psPos[]}} \defeq \left[
  \psfVec[\iKerCov]\paren{\x_{1} - \psPos[]}, \dots,
  \psfVec[\iKerCov]\paren{\x_{\nPx} - \psPos[]} \right]
  \ .
\end{equation*}

\paragraph{Identifiability of Gaussian mixture weights}
From \refeq{eq:psf:ifm-gm-vec} it is apparent that only the product between the PS
intensity and the GM weights is identifiable.
We define the \emph{net mixture intensity} as
$\wKerNetVec[\iKerCov] \defeq \psPhi[]\, \wKerNormedVec \in \Rpos^{\nKerPos}$: the
PS intensity is spread among the kernels in the dictionary as virtual
intensities. We re-parameterise the vector of expected photon count as
$\expPhCountVec[] \paren{\psPos[], \wKerNetVec, \params^\dict} =
\expPhCountVec[] \paren{\psPos[], \psPhi[], \params^\GM}$, where
$\wKerNetVec \in \Rpos^{\abs{\dict}}$ is the stacked vector of net mixture
intensities.

\subsubsection{Regularisation functional}
In general, the regularisation term $\regTerm$ in~\refeq{eq:psf:optim-general}
accounts for the prior knowledge about the PS position and intensity, and the
PSF model parameters.
However, in this paper we aim at showing how to explore the accuracy--efficiency
trade-off based on the GM model.
Therefore, we focus on regularising only the PSF model parameters, \ie
$\regTerm^\GM\paren{\params}$.

One way to trade accuracy for efficiency is to penalise the effective number of
Gaussian kernels, by penalising the size of the support of the net mixture
intensity vector:
\begin{equation*}
  \label{eq:psf:def-reg-mg}
  \regTerm^\GM \paren{\wKerNetVec} \defeq
  \nPx \sum_{\iKerCov \in \iKerCovSet} \frac{\reg_\iKerCov}{\reg} \abs*{\supp\paren*{\wKerNetVec[\iKerCov]}}\ ,
\end{equation*}
where we scale $\reg$ by the number of pixels to have a comparable
regularisation for different image data.
Solving problem~\refeq{eq:psf:optim-general} with this regularisation functional
aims precisely at finding the optimal trade-off between accuracy (minimising the
$\nll$) and efficiency (minimising the number of non-zero mixture weights).
The regularisation parameters are stacked into a vector, denoted
$\regVec \defeq \paren*{\reg_1,\dots,\reg_{\nKerCov}} \in \Rpos^\nKerCov$, and
allow exploring this trade-off: small values of $\abs{\regVec}$ favour accuracy
over efficiency, and large values \viceversa.

\subsubsection{Convex relaxation}
\label{sec:psf:convex-relaxation-mg}
Problem~\refeq{eq:psf:optim-general} has multiple sources of non-convexity for the
GM model: the joint estimation of PS position and intensity, the PSF parameters
estimation, and the cardinality regularisation functional for the GM model.
Nevertheless, it has also convex components: the constraint sets are convex, and
$\nll$ is convex in the mean photon count vector.
If the PS position $\psPos[]$ and the dictionary parameters $\params^\dict$ are
fixed, the remaining non-convexity is $\regTerm^\GM$.
To handle more tractable problems, we use a convex sparsity-inducing
regulariser, denoted $\regTerm^\GM_\CONVEX$, based on the popular $\ell_1$~norm
\cite{Bach2012}:
\begin{equation}
  \label{eq:psf:def-reg-mg-co}
  \regTerm^\GM_\CONVEX \paren{\wKerNetVec}
  \defeq \nPx \sum_{\iKerCov \in \iKerCovSet} \frac{\reg_\iKerCov}{\reg} \Lone{\wKerNetVec[\iKerCov]}\ .
\end{equation}

\subsubsection{Problem formulation for BW and SG PSF models}
For BW and SG, problem~\refeq{eq:psf:optim-general} amounts to solving the
maximum likelihood problem:
\begin{equation}
  \label{eq:psf:optim-sg-bw}
  \begin{aligned}
    \paren{\psPosHat[], \psPhiHat[], \paramsHat} \defeq & \argmin{\psPos[],\, \psPhi[],\, \params} \nll\paren{
      \psPos[], \psPhi[], \params \mid \obsGVMat[]
    } \\
    \st & \psPos[] \in \imageDom, \ \psPhi[] \in \Rpos,\ \params
    \in \paramSet \ .
  \end{aligned}
\end{equation}

\subsubsection{Problem formulation for GM PSF models}
We propose solving problem~\refeq{eq:psf:optim-general} in three steps.

\paragraph{PS position estimation, $\psPosHat[]$}
\label{sec:psf:gm:ps-position-estimation}
We solve the \emph{parametric blind PS deconvolution problem} for the SG model
given by \refeq{eq:psf:optim-sg-bw}.

\paragraph{Mixture weight support estimation}
We fix the PS position to its estimated value and solve the \emph{GM PSF
  calibration problem} using the regulariser~\refeq{eq:psf:def-reg-mg-co}:
\begin{equation}
    \label{eq:psf:optim-mg-l1}
    \wKerNetVecHat[\regVec] \defeq \argmin{\wKer \in \Rpos^{\abs{\dict}}} \nllNormed\paren{
      \wKer \mid \obsGVMat[], \psPosHat[], \params^\dict
    } + \sum_{\iKerCov \in \iKerCovSet} \reg_\iKerCov \Lone*{\wKer[\iKerCov]}  \ , 
\end{equation}
where $\nPx\, \nllNormed \defeq \nll$.
We estimate the support by thresholding the weights above~$\wKerMin$:
\begin{equation*}
  \widehat{\supp}\paren*{\wKerNetVecHat[\regVec]} \defeq
  \set*{(\iKerCov,\iKerPos) \in \iKerCovSet \iKerPosSet \mid \parenEl{\wKerNetVecHat[\regVec,\iKerCov]}_\iKerPos > \wKerMin} \ .
\end{equation*}
The \emph{effective dictionary} is defined by keeping the kernels in the
estimated support:
\begin{equation}
  \label{eq:psf:effective-dict}
  \dictPosHat_{\regVec} \defeq \set*{\psfFun^\SG_{\params_\iKerCov}(\cdot-\x_\iKerPos) \in \dict \mid (\iKerCov, \iKerPos) \in \widehat{\supp}\paren*{\wKerNetVecHat[\regVec]}}\ .
\end{equation}

\paragraph{Debiasing}
\label{sec:psf:gm:debiasing}
The convex problem \refeq{eq:psf:optim-mg-l1} brings a known problem due to
using the $\ell_1$ norm: a bias (see \eg \cite{Deledalle2015}) in the intensity
estimates $\wKerNetVecHat[\regVec]$.
We use a refitting strategy based on maximum likelihood to \emph{debias the
mixture weights}:
\begin{equation}
  \label{eq:psf:optim-mg-debiasing}
  \wKerNetVecDeb[\regVec] \defeq
  \argmin{\wKer \in \Rpos^{\abs{\dictPosHat_{\regVec}}}}
  \nll\paren*{\wKer \mid \obsGVMat[], \psPosHat[], \dictPosHat_{\regVec}} \ . 
\end{equation}
We compute the estimates of the PS intensity as $\psPhiHat[] \defeq
\Lone{\wKerNetVecDeb[\regVec]}$ and mixture weights as $\wKerNormedVecHat \defeq
\paren{\psPhiHat[]}^{-1}\wKerNetVecDeb[\regVec]$.

\subsection{Algorithms}
\label{sec:psf:algorithms}
%

\begin{figure}[!t]
  \removelatexerror
  \begin{algorithm}[H]
    \label{alg:asb-psf}
    \caption{Fully-split ASB for solving problem~\refeq{eq:psf:optim-mg-l1}}
    \SetKwInOut{Input}{Input}
    \SetKwInOut{Output}{Output}

    \Input{
        $\obsGVMat[]$, $\bgPhi[]$, $\regVec$, $\step$,
        $\set{\blurMat_\iKerCov, \b^0_\iKerCov, \w^0_\iKerCov}_{\iKerCov\in\iKerCovSet}$}
      \BlankLine
      \Output{ $\wKerNetVecHat[\regVec] \defeq \w_{3}^{\infty}$}
      \BlankLine
      $ \forall \iKerCov \in \iKerCovSet :
      %
      %
      \w^0_{1\iKerCov} = \obsPhCountVec[], \
      \w^0_{2\iKerCov} =
      \w^0_{3\iKerCov} = \obsPhCountVec[\iKerCov], \
      \b^0_\iKerCov =\zerosVec_{\nPx+2\nKerPos}$
      \BlankLine
      \While{\textsc{not converged}}{
      \For{$\iKerCov = 1$ \KwTo $\nKerCov$}{
        LS sub-problem:
        \begin{fleqn}[\dimexpr(\leftmargini-\labelsep)*1]
          \begin{equation}
            \label{eq:psf:alg-sp-ls}
            \wKerLSn = \argmin{\wKerLS} \Ltwo{\bc_\iKerCov +
              \operator_\iKerCov \wKerLS - \wc_\iKerCov}^2
          \end{equation}
        \end{fleqn}
        $\nll$ sub-problem:
        \begin{fleqn}[\dimexpr(\leftmargini-\labelsep)*1]
          \begin{align}
            \label{eq:psf:alg-sp-nll}
            \wn_{1\iKerCov} = \Prox_{\step\nllNormed}\paren*{\bc_{1\iKerCov} + \blurMat_\iKerCov\paren{\psPos[]} \wKerLSn}
          \end{align}
        \end{fleqn}
        %
        $\ell_1$--$\ell_2^2$ sub-problem:
        \begin{fleqn}[\dimexpr(\leftmargini-\labelsep)*1]
          \begin{gather}
            \label{eq:psf:alg-sp-lone}
            \wn_{2\iKerCov} = \Prox_{\step\reg_\iKerCov\Lone{\cdot}}\paren{\bc_{2\iKerCov} + \wKerLSn}
          \end{gather}
        \end{fleqn}
        Positivity constraint sub-problem:
        \begin{fleqn}[\dimexpr(\leftmargini-\labelsep)*1]
          \begin{gather}
            \label{eq:psf:alg-sp-proj}
            \wn_{3\iKerCov} = \Proj_{\Rpos^{\nKerPos}}\paren{\bc_{3\iKerCov} + \wKerLSn}
          \end{gather}
        \end{fleqn}
        Bregman update (dual gradient ascent):
        \begin{fleqn}[\dimexpr(\leftmargini-\labelsep)*1]
          \begin{equation}
            \bn_\iKerCov = \bc_\iKerCov + \operator_\iKerCov \wKerLSn - \wn_\iKerCov
          \end{equation}
        \end{fleqn}
      }
    }
  \end{algorithm}
\end{figure}


%
The main difference in solving the parametric blind PS deconvolution problem for
different PSF models is the dimensionality of the solution space.
For the SG and BW PSF models, problem~\refeq{eq:psf:optim-sg-bw} requires estimating
$6$ and $7$ parameters, respectively. Both are low-dimensional smooth non-convex
optimisation problems that can be solved by a general-purpose black-box
optimiser.
In contrast, the GM PSF calibration problem~\refeq{eq:psf:optim-mg-l1} is a
large-scale non-differentiable convex optimisation problem.
We derive an efficient and modular algorithm to solve~\refeq{eq:psf:optim-mg-l1}
based on the alternating split Bregman strategy, and discuss the algorithms for
the debiasing step.

\subsubsection{Estimating the support of MG PSF model}
\label{sec:psf:asb-solving}
We define the energy functional in problem~\refeq{eq:psf:optim-mg-l1} as:
%
%
\begin{equation*}
  \energy\paren{\wKer}
  \defeq \nllNormed\paren{\wKer \mid \obsGVMat[], \psPosHat[], \params^\dict} +
  \sum_{\iKerCov \in \iKerCovSet} \reg_\iKerCov \Lone*{\wKer[\iKerCov]}
  + \indicatorOptim_\mathcal{S} \paren{\wKer}\ ,
\end{equation*}
where $\indicatorOptim_\mathcal{S}\paren{\wKer}$ with $\mathcal{S}
\defeq{{\Rpos^{\mathsmaller{\abs{\dict}}}}}$ is the indicator functional that
assumes $0$ if $\wKer$ is componentwise non-negative and $+\infty$ otherwise.
This problem has an additive structure at two levels.
First, the functional is the sum of three terms: the $\nll$, the $\ell_1$
regulariser, and the indicator function $\indicatorOptim_\mathcal{S}
\paren{\wKer}$.
The second level of additivity comes from the sum over kernels in the mean
vector~\refeq{eq:psf:ifm-gm-vec}.
We exploit this additivity structure and derive the algorithm in three steps:
%

\paragraph{Operator splitting}
\label{sec:psf:op-splitting}
We exploit the first level of additivity by introducing dummy variables to split
the different terms: $\w_\iKerCov = \transpose{ \begin{bmatrix}
\transpose{\w_{1\iKerCov}}\ \transpose{\w_{2\iKerCov}}\
\transpose{\w_{3\iKerCov}} \end{bmatrix}} \defeq \operator_\iKerCov
\wKerNetVec[\iKerCov] \in \R^{\nPx+2\nKerPos} $
%
%
with $ \operator_\iKerCov \defeq \transpose{ \begin{bmatrix}
\transpose{\blurMat}_\iKerCov\paren{\psPosHat[]}\ \eyeMat_{\nKerPos}\
\eyeMat_{\nKerPos} \end{bmatrix}} \in \R^{\paren{\nPx+2\nKerPos} \times
\nKerPos} \ , $
%
%
where $\eyeMat_{\nKerPos} \in \R^{\nKerPos \times \nKerPos}$ is the identity matrix.
We define the stacked vector $\w \in \R^{\nKerCov\nPx+2\abs{\dict}}$ and the
block-diagonal matrix
$\operator \defeq \diag\paren{\operator_1, \dots, \operator_{\nKerCov}} \in
\R^{\paren{\nKerCov\nPx+2\abs{\dict}} \times \abs{\dict}}$, such that
$\w = \operator\wKerNetVec$.
In addition, we denote the subset of dummy variables for the operator
$o \in \set{1,\, 2,\, 3}$ stacked into a vector as
$\w_{o:} \defeq
\vectorise\paren*{\set{\w_{o\iKerCov}}_{\iKerCov\in\iKerCovSet}}$.
We can rewrite energy $\energy\paren{\wKer}$ to highlight its additive
structure: $ \energy\paren{\w} \defeq \energy\paren{\w_{1:}, \w_{2:}, \w_{3:}}
\defeq \nllNormed(\w_{1:}) + \sum_{\iKerCov \in \iKerCovSet} \reg_\iKerCov
\Lone*{\w_{2\iKerCov}} + \indicatorOptim_\mathcal{S} \paren{\w_{3\iKerCov}}$,
%
%
where for simplifying the notations, we implicitly assume the conditional
dependence of $\nllNormed$ on the image data $\obsGVMat[]$, the estimated point
source position $\psPosHat[]$, and the dictionary parameters $\params^\dict$.

\paragraph{Fully-split Bregman}
The next step is to fully split $\w$ from $\wKerNetVec$ by applying split
Bregman to the equivalent problem $\argmin{\wKer,\w} \inner{\zerosVec_{\abs{\dict}},\,\wKer} +
\energy\paren{\w}$,
%
%
writing at iteration $\iIter+1$:
\begin{align*}
  \paren{\wKern, \wn} & = \argmin{\wKer, \w} \inner{\zerosVec_{\abs{\dict}},\,\wKer} + \Psi^{\iIter}\paren{\wKer, \w}\\
  \bn & = \bc + \operator \wKern - \wn\ ,
\end{align*}
%
%
where $\Psi^{\iIter}\paren{\wKer, \w} \defeq \energy\paren{\w} +
\paren{2\step}^{-1} \Ltwo{\bc + \operator \wKer - \w}^2$, $\b~\in~\R^{\nKerCov\nPx+2\abs{\dict}}$ is the vector of the Bregman dual variables
stacked similarly to $\w$, and $\step$ is the dual-ascent step-size.

\paragraph{Gauss-Seidel-like alternation}
We exploit the second level of additivity by splitting over the kernels the
optimisation problem in the Bregman iteration.
The algorithm at iteration $\iIter$ for kernel $\iKerCov$ reads:
\begin{align}
  \wKerLSn & = \argmin{\wKerLS} \inner{\zerosVec_{\abs{\dict}}, \wKerLS} + \frac{1}{2\step} \Ltwo{\bc_\iKerCov + \operator_\iKerCov \wKerLS - \wc_\iKerCov}^2 \label{eq:psf:sp-ls}\\
  \wn_\iKerCov & = \argmin{\w_\iKerCov} \energy_\iKerCov^\iIter\paren{\w_\iKerCov} + \frac{1}{2\step} \Ltwo{\bc_\iKerCov + \operator_\iKerCov \wKerLSn - \w_\iKerCov}^2\notag\\
  & \eqdef \Prox_{\step\energy_\iKerCov^\iIter} \paren*{\bc_\iKerCov + \operator_\iKerCov \wKerLSn}\label{eq:psf:sp-prox}\\
  \bn_\iKerCov & = \bc_\iKerCov + \operator_\iKerCov \wKerLSn-\wn_\iKerCov \label{eq:psf:sp-bregman}\ .
\end{align}
where the energy $\energy_\iKerCov^\iIter\paren{\w_\iKerCov}$ introduces the
coupling between kernels and relies on the set of $\iKerCov - 1$ weights that
are already updated at the $\iIter$-th iteration:
$\energy_\iKerCov^\iIter\paren{\w_\iKerCov} \defeq \nllNormed\paren{...,
\wn_{1\iKerCov-1}, \w_{1\iKerCov}, \wc_{1\iKerCov+1}, ...} + \reg_\iKerCov
\Lone*{\w_{2\iKerCov}} + \indicatorOptim_\mathcal{S}\paren{\wKer[3\iKerCov]}$.
%

\paragraph{Sub-problems solution}
\label{sec:psf:solution:sub-problems}
The fully-split ASB strategy amounts to solving three standard problems: a
least-squares (LS) problem~\refeq{eq:psf:sp-ls}, a proximal map
evaluation~\refeq{eq:psf:sp-prox}, and a linear
update~\refeq{eq:psf:sp-bregman}. Further, it leads to a decomposition of the
proximal map~\refeq{eq:psf:sp-prox} in three independent proximal
maps~\refeq{eq:psf:alg-sp-nll}, \refeq{eq:psf:alg-sp-lone}, and
\refeq{eq:psf:alg-sp-proj}, that decouple across pixels.
The algorithm is shown in \refalg{alg:asb-psf}, where $\obsPhCountVec[\iKerCov]$
denotes the image data at the virtual source positions.

The solution to the LS sub-problem~\refeq{eq:psf:sp-ls} is straightforward:
\begin{equation}
  \label{eq:psf:sp-ls-solution}
  \wn_\iKerCov =
  \inverse{\transpose{\operator_\iKerCov}\operator_\iKerCov}
  \transpose{\operator_\iKerCov}\paren*{\wc_\iKerCov - \bc_\iKerCov}\ ,
\end{equation}
with $\transpose{\operator_\iKerCov}\operator_\iKerCov = 2\eyeMat_{\nKerPos} +
\transpose{\blurMat_\iKerCov}\blurMat_\iKerCov$ and
$\transpose{\operator_\iKerCov}\paren*{\wc_\iKerCov - \bc_\iKerCov} =
\transpose{\blurMat_\iKerCov}\paren{\wc_{1\iKerCov} - \bc_\iKerCov} +
\wc_{2\iKerCov} - \bc_{2\iKerCov} + \wc_{3\iKerCov} - \bc_{3\iKerCov}$,
%
%
where $\blurMat_\iKerCov \defeq \blurMat_\iKerCov\paren{\psPos[]}$.
The inverse operator can be pre-computed outside the main iteration. Moreover,
digital dictionaries can be efficiently evaluated using a spectral solver based
on the FFT.

The proximal maps \refeq{eq:psf:alg-sp-lone} and \refeq{eq:psf:alg-sp-proj} are also
known:
\refeq{eq:psf:alg-sp-lone} is a componentwise soft-thresholding of
$\bc_{2\iKerCov} + \wKerLSn$ with threshold $\step\reg_\iKerCov$,
and \refeq{eq:psf:alg-sp-proj} is a componentwise projection of
$\bc_{3\iKerCov} + \wKerLSn$ on $\Rpos$.
The solution to the proximal~\refeq{eq:psf:alg-sp-nll} is derived
in~\cite{Setzer2010,Paul2013}.
However, due to the additive structure of the mean vector and the
Gauss-Seidel-like alternation, we provide more details for this proximal.

At iteration $\iIter$, when the alternation reaches kernel $\iKerCov$, the mean
vector~\refeq{eq:psf:ifm-gm-vec} can be rewritten in terms of the dummy variables:
\begin{equation*}
  \expPhCountVec[\iKerCov]^{\iIter+1} \defeq
  \sum_{\mathclap{\iKerCovTmp=1}}^{\iKerCov-1} \wn_{1\iKerCovTmp} +
  \w_{1\iKerCov} +
  \sum_{\mathclap{\iKerCovTmp=\iKerCov+1}}^{\nKerCov}\wc_{1\iKerCovTmp} +
  \bgPhi[] \onesVec_\nPx
  \eqdef \w_{1\iKerCov} + \rn_\iKerCov\ .
\end{equation*}
Instead of solving \refeq{eq:psf:alg-sp-nll} in terms of $\w_{1\iKerCov}$, we solve
the problem in terms of $\expPhCountVec[\iKerCov]$ and update $\wn_{1\iKerCov}$
by subtracting the residuals $\rn_\iKerCov$:
\begin{align}
  \expPhCountVec[\iKerCov]^{\iIter+1} &= \Prox_{\step\nllNormed}\paren*{\bc_{1\iKerCov} + \blurMat_\iKerCov \wKerLSn + \rn_\iKerCov} \label{eq:psf:alg-sp-nll-mean} \\
  \wn_{1\iKerCov}  &= \expPhCountVec[\iKerCov]^{\iIter+1} - \rn_\iKerCov \ .
\end{align}
Evaluating~\refeq{eq:psf:alg-sp-nll-mean} amounts to choosing the positive solution
of the following quadratic equation in $\parenEl{\expPhCountVec[\iKerCov]}_\iPx$
defined for the $\iPx$-th pixel:
\begin{equation*}
  \parenEl{\expPhCountVec[\iKerCov]}_\iPx^2 + \paren*{ \frac{\step}{\nPx} - \transpose{\indicatorVec_\iPx} \vec{c}^{\iIter}_{\iKerCov} } - \frac{\step}{\nPx} \parenEl{\obsPhCountVec}_\iPx = 0\ ,
\end{equation*}
with $\vec{c}^{\iIter}_{\iKerCov} \defeq \bc_\iKerCov + \blurMat_\iKerCov
\wKerLSn + \rn_\iKerCov$.

\subsubsection{Debiasing mixture weights of MG PSF model}
\label{sec:psf:solution:debiasing}
The gradient of $\nll$ with respect to the mixture weights can be derived
analytically:
\begin{equation*}
  \label{eq:psf:nll-gradient}
  \partial_{\wKer[\iKerCov]} \nll\paren{\wKer} =
  \sum_{\iPx = 1}^{\nPx} \paren*{1 - \frac{\parenEl{\obsPhCountVec}_\iPx}{\parenEl{\expPhCountVec[]\paren{\psPos[], \wKer}}_\iPx}}\transpose{\blurMat_\iKerCov} \indicatorVec_\iPx\ .
\end{equation*}
To solve~\refeq{eq:psf:optim-mg-debiasing}, we use gradient-based algorithms
allowing positivity constrains.




\section{Experiments}
\label{sec:psf:exp}
%

\subsection{Illustrative example: synthetic 1-D}
\label{sec:psf:exp-1d-ex}


\begin{table*}[!t]
  \renewcommand{\arraystretch}{1.3}
  \newcommand{\head}[1]{\textnormal{\textbf{#1}}}
  \newcommand{\normal}[1]{\multicolumn{1}{c}{#1}}
  \newcommand{\na}{\emph{n.a.}}

  \colorlet{tableheadcolor}{gray!25} 
  \newcommand{\headcol}{\rowcolor{tableheadcolor}} %
  \colorlet{tablerowcolor}{gray!10} 
  \newcommand{\rowcol}{\rowcolor{tablerowcolor}} %
  \setlength{\aboverulesep}{0pt}
  \setlength{\belowrulesep}{0pt}

  \newcommand*{\rulefiller}{
    \arrayrulecolor{tableheadcolor}
    \specialrule{\heavyrulewidth}{0pt}{-\heavyrulewidth}
    \arrayrulecolor{black}}

  \caption{Parameters of the image formation process}
  \label{table:psf:imaging-parameters}
  \centering
  \begin{tabular}{c*{12}{c}}
    \toprule
    \headcol \head{Dataset} & \multicolumn{3}{c}{\head{Object-to-pixel}} & \multicolumn{4}{c}{\head{Sampling}} & \multicolumn{4}{c}{\head{Pixel-to-image}} & \head{Object}\\
    \rulefiller \cmidrule(lr){2-4} \cmidrule(lr){5-8} \cmidrule(lr){9-12} \cmidrule(lr){13-13}
    \headcol & $\wavelength$ [\SI{}{\nm}] & $\NA$ & $\refIndImm$ &  $\nHeight\times\nWidth\times\nSlices$ & $\lateralSampling$ [\SI{}{\nm}] & $\axialSampling$ [\SI{}{\nm}] & $\tE$ [\SI{}{\ms}] & $\quantEfficiency$ & $\mGain$ & $\ADU$ & $\cameraBias$ & diameter [\SI{}{\nm}]\\
    \midrule
    Synthetic 1-D & $474$ & $1.45$ & $1.518$ & $1 \times 101 \times 1$ & $50$ & $-$ & $1$ & $1$ & $1$   & $1$     & $0$ & $\delta$\\
    \midrule
    SBW & $474$ & $1.45$ & $1.518$ & $81 \times 81 \times 101$ & $65$ & $200$ & $21$ & $0.81$ & $1$   & $2.0$     & $100$ & $\delta$\\
    WFFM & $620$ & $1.45$ & \na     & $81 \times 81 \times  31$ & $65$ & $200$ & $21$  & $0.81$ & $1$ & $2.14$ & $98.24$ & $50$\\
    LSCM & $520$ & $1.45$ & \na     & $61 \times 61 \times 101$ & $133$ & $50$ & $12$  & $0.70$ & $200$ & $6.44$  & $398.06$ & $100$\\
    \bottomrule
  \end{tabular}
\end{table*}


\begin{figure*}[!t]
  \centering
  \includegraphics[width=0.9\textwidth]{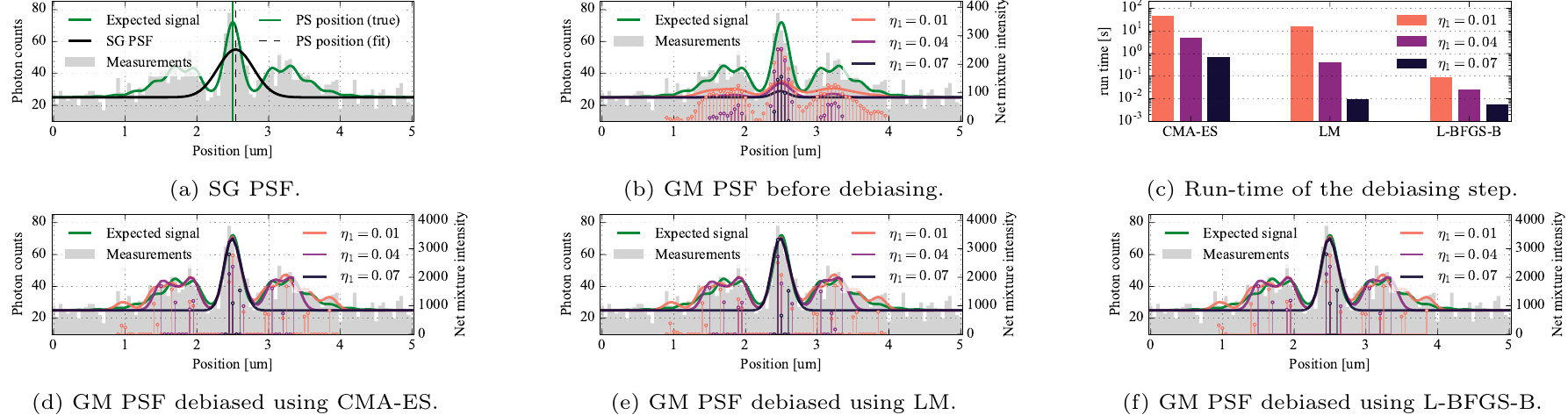}
  \subfloat{\label{fig:psf:1d-deb-cmaes}}
  \subfloat{\label{fig:psf:1d-deb-asb}}
  \subfloat{\label{fig:psf:1d-deb-run_time}}
  \subfloat{\label{fig:psf:1d-deb-LM}}
  \subfloat{\label{fig:psf:1d-deb-CMAES}}
  \subfloat{\label{fig:psf:1d-deb-LBFGSB}}
  \caption{ %
    \textbf{SG PSF model and GM PSF model with a one-kernel dictionary for a
      one-dimensional synthetic signal.}
    \refsubfig{fig:psf:1d-deb-cmaes} Parametric blind PS deconvolution for the SG
    model.
    \refsubfig{fig:psf:1d-deb-asb} GM model (continuous line) and mixture weights
    support (weighted Dirac comb shown as vertical bars) estimated using
    \refalg{alg:asb-psf} for different amount of regularisation.
    \refsubfig{fig:psf:1d-deb-run_time} Run-time of debiasing using the
    covariance matrix evolution evolution strategy (CMA-ES), the Levenberg-Marquardt
    (LM) and the limited-memory Broyden–Fletcher–Goldfarb–Shanno algorithm with box
    constraints (L-BFGS-B).
    \refsubfig{fig:psf:1d-deb-LM} - \refsubfig{fig:psf:1d-deb-LBFGSB} GM model after
    debiasing. }
\label{fig:psf:1d-deb}
\end{figure*}


\subsubsection{Imaging settings}
We place an idealised PS object at
$\psPos[] = (\SI{2.5}{\um}, \SI{0}{\um}, \SI{1.5}{\um})$ in the 3D imaging
volume.
We model the optical distortions using the BW PSF.
Using the forward problem in \refsection{sec:psf:fwd}, we model the signal acquired
by a one-dimensional camera detector along the x-axis in the range
$x \in \intervalcc{\SI{0}{\um}}{\SI{5}{\um}}$, at $y = z = \SI{0}{\um}$
(the imaging parameters are summarised in \reftable{table:psf:imaging-parameters}).
This one-dimensional synthetic signal captures realistic features found in
experimental PSF image data: a prominent central mode with side lobes that are
characterised by a rich structure and an amplitude significantly above the
background signal.
%
%
In what follows, we assume the background intensity known.

\subsubsection{SG model}

We solve the parametric blind PS deconvolution problem using the covariance
matrix adaptation evolutionary strategy (CMA-ES) algorithm \cite{Hansen2003}. It
is a popular and well-tested derivative-free algorithm designed to solve
low-to-moderate dimensional non-convex optimisation problems
(\reffig{fig:psf:1d-deb-cmaes}).
We observe that the estimated PS position ($\hat{x} = \SI{2.539}{\um}$) is close
to the true value.
However, the approximation accuracy of the central mode is poor: its width is
overestimated resulting in an underestimated intensity.

\subsubsection{GM model with a one-kernel dictionary}
We use a digital dictionary, denoted $\dict_1^{1\mathrm{d}}$, made of Gaussian
kernels of standard deviation $\SI{0.1}{\um}$, placed at every pixel centres,
\ie $\params^{\dict_1^{1\mathrm{d}}} = \set{0.1} \times
\set{\x_{\iPx}}_{\iPx\in\iPxSet}$.
We estimate the net mixture intensities by applying \refalg{alg:asb-psf} for
different values of the regularisation parameter and build the GM model by using
a threshold of $\wKerMin = 0.1$ in \refeq{eq:psf:effective-dict}.
The results are shown on \reffig{fig:psf:1d-deb-asb}.
For smaller regularisation, the estimated
support of the GM model covers the support of the true signal.
For higher regularisation, the estimated support overlaps only with the regions
of high signal: the central mode and side lobes for $\reg_1=0.04$, only the
central mode for $\reg_1=0.07$.
However, we observe that the estimated intensity is biased, as expected (see
\refsection{sec:psf:gm:debiasing}).

We solve problem~\refeq{eq:psf:optim-mg-debiasing} using CMA-ES and two
gradient-based algorithms that allow enforcing positivity constraints:
Levenberg-Marquardt (LM), and the limited-memory
Broyden–Fletcher–Goldfarb–Shanno algorithm with box constraints (L-BFGS-B).
The three algorithms result in similar reconstructions, but L-BFGS-B has the
lowest run-time (\reffig{fig:psf:1d-deb}).
We observe that the debiasing step induces even more sparsity.

The estimated GM model improves compared to SG model in two ways:
the width and intensity of the central lobe are accurately reconstructed, even
for a high regularisation, \ie when the PSF is modelled with only a few kernels;
for smaller regularisation, the GM model also captures the side lobes.

\subsubsection{GM model with a two-kernel dictionary}
%

\begin{figure*}[!t]
  \centering
  \includegraphics[width=0.9\textwidth]{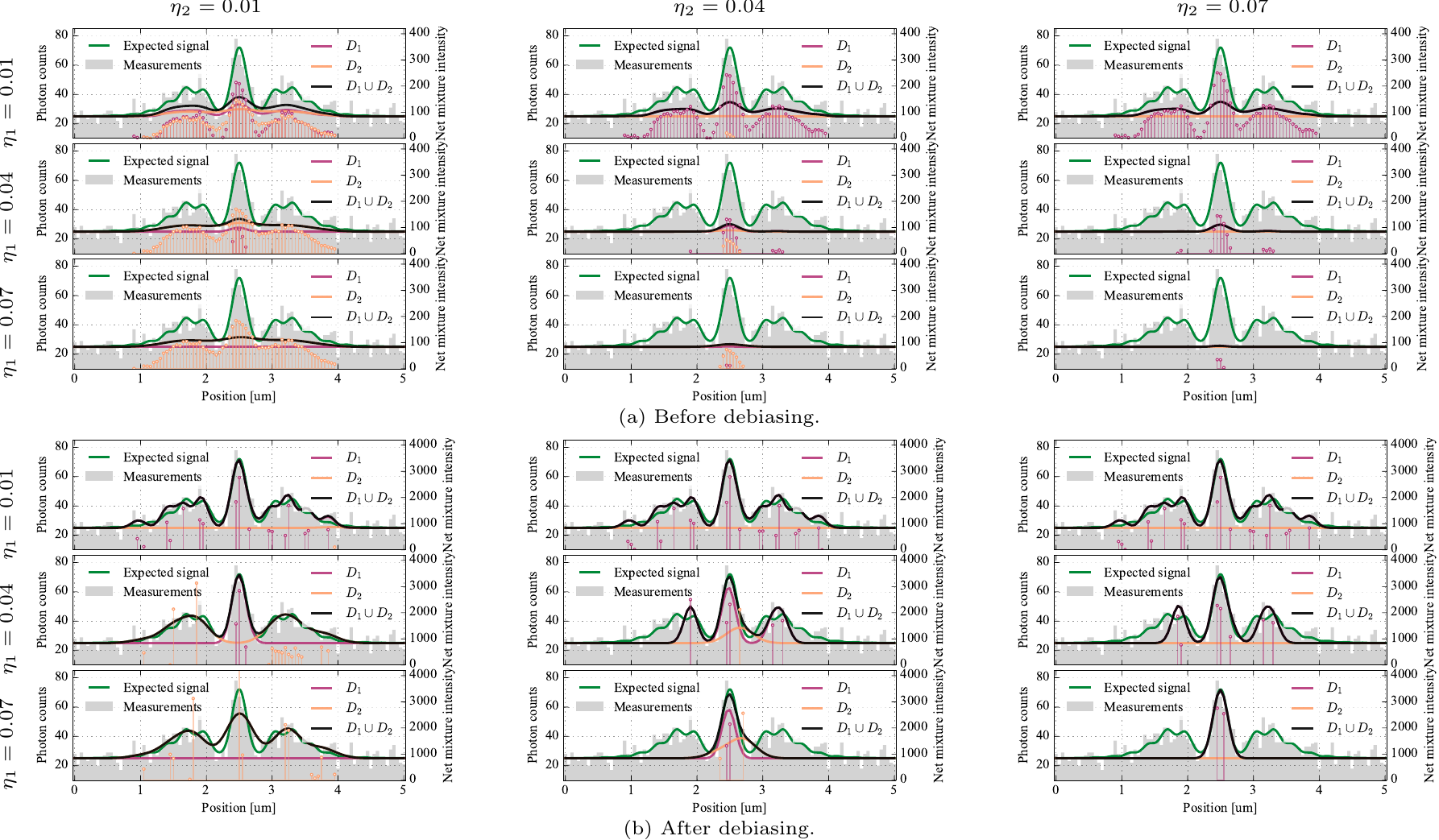}
  \subfloat{\label{fig:psf:1d-multi-before}}
  \subfloat{\label{fig:psf:1d-multi-after}}
  \caption{
    \textbf{GM PSF model with a two-kernel dictionary for a one-dimensional
      synthetic signal.}
    The GM model (continuous line) and the underlying mixture weights (weighted
    Dirac comb shown as vertical bars) are estimated using \refalg{alg:asb-psf} for
    different amount of regularisation before \refsubfig{fig:psf:1d-multi-before} and
    after \refsubfig{fig:psf:1d-multi-after} debiasing.
    The GM  models are evaluated separately for the smaller kernel (purple), the
    bigger kernel (orange), and both kernels (black).
  }
  \label{fig:psf:1d-multi}
\end{figure*}


%
We add to the previous dictionary a second kernel of standard deviation
$\SI{0.2}{\um}$ such that the new dictionary, denoted
$\dict_2^{1\mathrm{d}}$, has parameters $\params^{\dict_2^{1\mathrm{d}}} =
\params^{\dict_1^{1\mathrm{d}}} \cup \set{0.2} \times
\set{\x_{\iPx}}_{\iPx\in\iPxSet}$.
Similarly to the one-kernel dictionary, we observe that the amount of
regularisation controls the number of each kernel in the effective dictionary
(\reffig{fig:psf:1d-multi-before}).
If the regularisation is equally increased for both kernels (plots on the
diagonal), the support size decreases. For a higher regularisation
($\reg_1=\reg_2=0.07$), only the smaller kernel contributes to modelling the
PSF.
If the regularisation parameter of one kernel changes, the cardinality of the
support of the other kernel remains the same.
At the debiasing step, if the regularisation parameter for both kernels are
equal, the preference is given to the smaller kernel as it allows capturing
finer details of the PSF (\reffig{fig:psf:1d-multi-after}, plots on the diagonal).
When we regularise more the smaller kernel, the bigger kernel is used to
approximate the wider lobes, whereas the smaller kernel is used to approximate
the narrow central mode (\reffig{fig:psf:1d-multi-after}, $\reg_1=0.04$,
$\reg_2=0.01$).

\subsection{GM PSF model: exploring accuracy--efficiency trade-off}
%

\begin{table}[!t]
  \renewcommand{\arraystretch}{1.3}
  \newcommand{\head}[1]{\textnormal{\textbf{#1}}}
  \newcommand{\normal}[1]{\multicolumn{1}{c}{#1}}
  \newcommand{\na}{\emph{n.a.}}

  \colorlet{tableheadcolor}{gray!25} 
  \newcommand{\headcol}{\rowcolor{tableheadcolor}} %
  \colorlet{tablerowcolor}{gray!10} 
  \newcommand{\rowcol}{\rowcolor{tablerowcolor}} %
  \setlength{\aboverulesep}{0pt}
  \setlength{\belowrulesep}{0pt}

  \newcommand*{\rulefiller}{
    \arrayrulecolor{tableheadcolor}
    \specialrule{\heavyrulewidth}{0pt}{-\heavyrulewidth}
    \arrayrulecolor{black}}

  \caption{Estimated and theoretical SG PSF models comparison}
  \label{table:psf:sg}
  \centering
  \begin{tabular}{c*{4}{c}}
    \toprule
    \headcol \head{Dataset} & \multicolumn{2}{c}{\head{Estimated parameters}} & \multicolumn{2}{c}{\head{Theoretical parameters}} \\
    \rulefiller \cmidrule(lr){2-3} \cmidrule(lr){4-5}
    \headcol & $\paren{\stdXY,\stdZ}$ [\SI{}{\nm}] & $\nll$ & $\paren{\stdXY,\stdZ}$ [\SI{}{\nm}] & $\nll$ \\
    \midrule
    SBW & $\paren{262,842}$ & $3.82\expe{+5}$ & $\paren{74,267}$ & $5.44\expe{+5}$ \\
    WFFM & $\paren{400,801}$ & $3.78\expe{+5}$ & $\paren{96,349}$ & $5.34\expe{+5}$ \\
    LSCM & $\paren{247,512}$ & $5.50\expe{+5}$ & $\paren{63,229}$ & $17.2\expe{+5}$ \\
    \bottomrule
  \end{tabular}
\end{table}


\begin{table}[!t]
  \renewcommand{\arraystretch}{1.3}
  \newcommand{\head}[1]{\textnormal{\textbf{#1}}}
  \newcommand{\normal}[1]{\multicolumn{1}{c}{#1}}
  \newcommand{\na}{\emph{n.a.}}

  \colorlet{tableheadcolor}{gray!25} 
  \newcommand{\headcol}{\rowcolor{tableheadcolor}} %
  \colorlet{tablerowcolor}{gray!10} 
  \newcommand{\rowcol}{\rowcolor{tablerowcolor}} %
  \setlength{\aboverulesep}{0pt}
  \setlength{\belowrulesep}{0pt}

  \newcommand*{\rulefiller}{
    \arrayrulecolor{tableheadcolor}
    \specialrule{\heavyrulewidth}{0pt}{-\heavyrulewidth}
    \arrayrulecolor{black}}

  \caption{Designs of dictionaries with a single kernel}
  \label{table:psf:dict-designs}
  \centering
  \begin{tabular}{c*{4}{c}}
    \toprule
    \headcol \head{Dataset} & \multicolumn{4}{c}{\head{Parameters $\paren{\stdXY, \stdZ}$ \textnormal{[\SI{}{\nm}]}}} \\
    \rulefiller \cmidrule(lr){2-5}
    \headcol & $\dict_1$ & $\dict_2$ & $\dict_3$ & $\dict_4$ \\
    \midrule
    SBW & $(262,842)$ & $(131,421)$ & $(87,281)$ & $(66,210)$ \\
    WFFM & $(400,801)$ & $(200,401)$ & $(133,267)$ & $(100,200)$ \\
    LSCM & $(246,509)$ & $(123,255)$ & $-$ & $-$ \\
    \bottomrule
  \end{tabular}
\end{table}

%
The one-dimensional synthetic example illustrates how the GM model improves the
model accuracy compared to SG, while controlling the model complexity by
designing the dictionary structure (\ie one/two kernels) and the amount of
regularisation.
We now apply our framework to three-dimensional image data: a synthetic image
stack generated by a BW model, and real fluorescent bead measurements acquired
by a widefield and a laser scanning confocal microscope.

\subsubsection{Imaging settings}
\label{sec:psf:imaging-settings}
The parameter of the image formation process are summarised in
\reftable{table:psf:imaging-parameters}.

\paragraph{Synthetic Born and Wolf (SBW)}
\label{sec:psf:imaging-settings:bw}
We use the VM framework~\cite{Samuylov2015} and the forward problem in
\refsection{sec:psf:point-source-object} to simulate an image stack of an
idealised PS object.
We simulate an idealised microscope governed by a BW PSF model, resulting in the
image stack shown in \reffig{fig:psf:psf-gallery:SBW}.

\paragraph{Widefied fluorescence microscopy (WFFM)}
SPHERO\textsuperscript{TM} fluorescent beads of \SI{50}{nm} in diameter
(excitation wavelength: \SI{576}{nm}) were imaged using a widefield setup
equipped with a CMOS camera ORCA-Flash 4.0 V2 (Hamamatsu).
We manually selected a region containing one bead, resulting in the image stack
in \reffig{fig:psf:psf-gallery:WFFM}.

\paragraph{Laser scanning confocal microscopy (LSCM)}
Tetraspeck\textsuperscript{TM} fluorescent beads of \SI{100}{nm} in diameter
(excitation wavelength: \SI{488}{nm}) were acquired using a confocal setup
equipped with an alpha Plan-Apochromat 100x/1.46 Oil DIC M27 objective (Carl
Zeiss) and an Evolve 512 Delta EM-CCD camera (Photometrics).
We manually selected a region containing one bead, resulting in the image stack
in \reffig{fig:psf:psf-gallery:LSCM}.

\subsubsection{Estimating the background intensity}

We estimate the background intensity by computing the median of the raw photon
count $\obsPhCountVec$ and divide it by the integration volume $c$.

\subsubsection{SG  model}
\label{sec:psf:tradeoff:model:sg}
We solve the parametric blind PS deconvolution problem using CMA-ES (SG).
We compare the estimated standard deviations with the theoretical values of the
SG model derived in \cite{Zhang2007} (SGT):
the parameters are overestimated due to the rich structure of the side lobes
(similar to the effect observed in the 1D synthetic example in
\reffig{fig:psf:1d-deb-cmaes}) but results in a lower $\nll$
(\reftable{table:psf:sg}).
The two models are show on \reffig{fig:psf:examples}.

\subsubsection{GM model with a one-kernel dictionary}
\label{sec:psf:tradeoff:model:mg:1k}
%

\begin{figure*}
  \centering
  \includegraphics[width=0.9\textwidth]{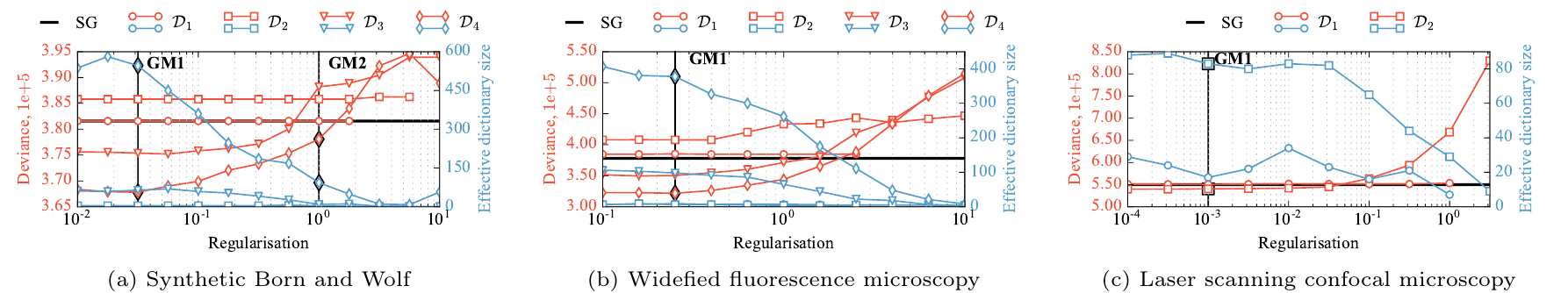}
  \subfloat{\label{fig:psf:tradeoff:one-kernel:bw}}
  \subfloat{\label{fig:psf:tradeoff:one-kernel:wf}}
  \subfloat{\label{fig:psf:tradeoff:one-kernel:lscm}}
  \caption{
  \textbf{Trade-off between accuracy and efficiency for the GM model with
  one-kernel dictionaries.}
  Results shown for the three datasets in \reffig{fig:psf:psf-gallery} and the
  digital dictionaries for different kernel size (SBW and WFFM: $\dict_1$,
  $\dict_2$, $\dict_3$, $\dict_4$; LSCM: $\dict_1$, $\dict_2$, see
  \reftable{table:psf:dict-designs}).
  For each dictionary, we explore the trade-off between \emph{accuracy}
  (measured as the deviance) and \emph{efficiency} (measured as the effective
  dictionary size) by varying the \emph{regularisation parameter}: the higher the
  regularisation, the lower the size, the higher the efficiency, the lower the
  accuracy.
  The horizontal black line indicates the deviance of the SG model estimated by
  solving the parametric blind PS deconvolution problem (see
  \refsection{sec:psf:tradeoff:model:sg}).
  We show examples of models referenced as \textbf{GM(.)}
  on~\reffig{fig:psf:examples}.
}
  \label{fig:psf:tradeoff:one-kernel}
\end{figure*}



\begin{figure*}
  \centering
  \includegraphics[width=0.9\textwidth]{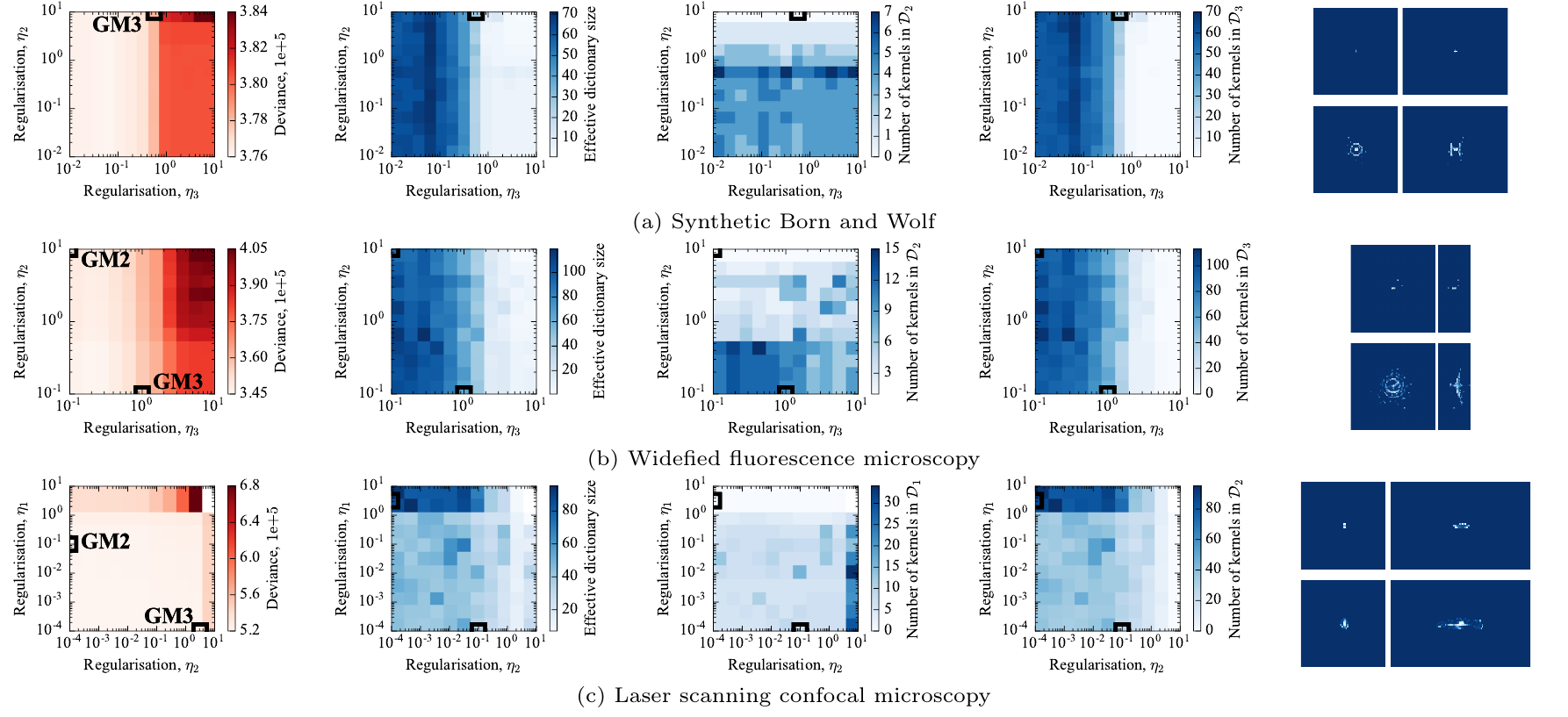}
  \subfloat{\label{fig:psf:tradeoff:two-kernels:bw}}
  \subfloat{\label{fig:psf:tradeoff:two-kernels:wf}}
  \subfloat{\label{fig:psf:tradeoff:two-kernels:lscm}}
  \caption{
  \textbf{Trade-off between accuracy and efficiency for the GM model with two-kernel dictionaries.}
  Results shown for the three datasets in \reffig{fig:psf:psf-gallery} and the
  digital dictionaries constructed from the one-kernel dictionaries used in
  \reffig{fig:psf:tradeoff:one-kernel} (SBW and WFFM: $\dict_2 \cup \dict_3$;
  CLSM: $\dict_1 \cup \dict_2$).
  We vary the amount of regularisation of each kernel in the dictionary to
  explore the trade-off between the \emph{accuracy} (column 1) and the
  \emph{efficiency} (column 2).
  For each amount of regularisation, we also show the number of larger (column
  3) and smaller (column 4) kernels in the effective dictionary.
  We compute the probability at each pixel of selecting a larger (column 5, top)
  or a smaller (column 5, bottom) kernel across all regularisation parameters.
  The plots display orthogonal z- and x- maximum projections of the
  probabilities.
  We show examples of models referenced as \textbf{GM(.)}
  on~\reffig{fig:psf:examples}.
}
  \label{fig:psf:tradeoff:two-kernels}
\end{figure*}



\begin{figure*}
  \centering
  \includegraphics[width=0.9\textwidth]{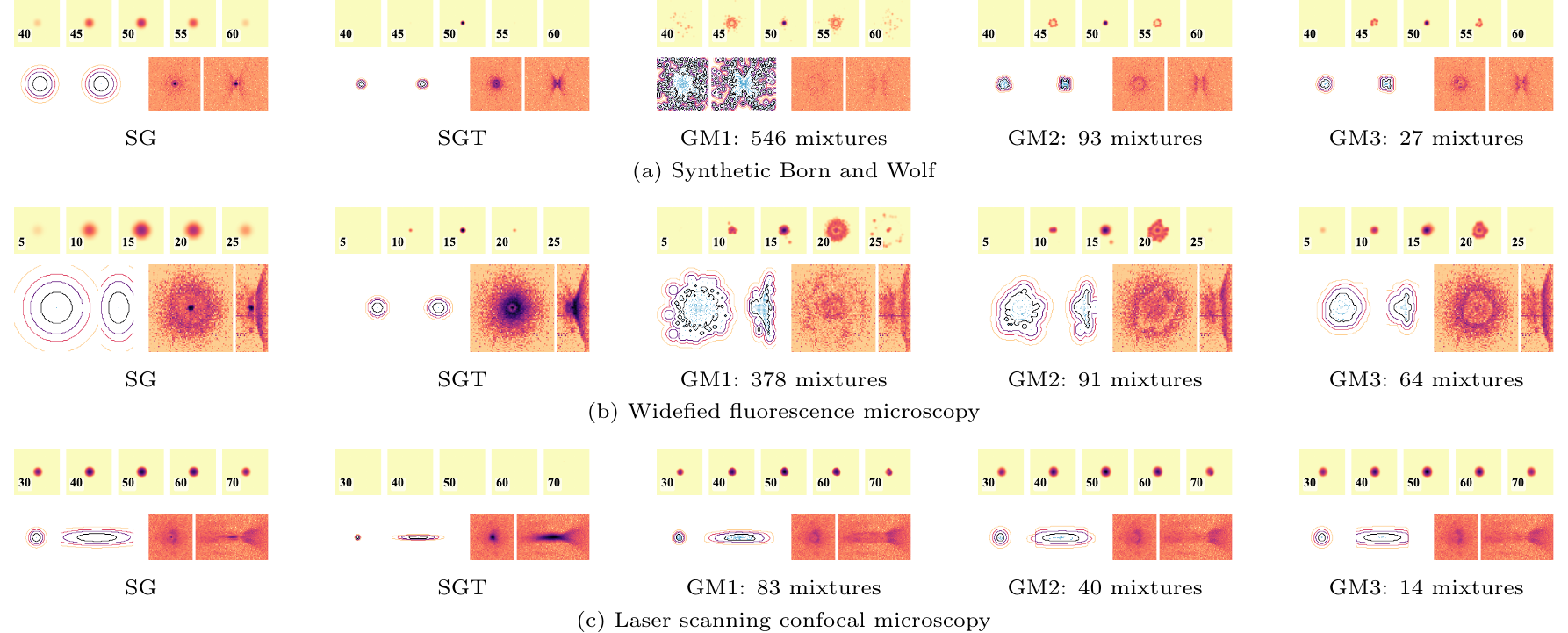}
  \caption{
  \textbf{Examples of SG, SGT and GM PSF models.}
  PSF models reconstructed from the datasets shown in
  \reffig{fig:psf:psf-gallery}: empirical SG model parametrised by solving the
  blind PS deconvolution (SG), theoretical SG model (SGT) and three GM models
  (GM1, GM2, GM3) selected as shown in \reffig{fig:psf:tradeoff:one-kernel} and
  \reffig{fig:psf:tradeoff:two-kernels}.
  For each PSF model, the panel is organised as follows.
  \emph{First row:} selected image slices corresponding to the same focal
  planes as in \reffig{fig:psf:psf-gallery}.
  \emph{Second row left:} highest density region (HDR) plot projected
  orthogonally along z- and x-axis. The HDR plots display the following
  quantiles for the isolines: $0.0$, $0.5$, $0.75$, $0.95$.
  \emph{Second row right:} orthogonal z- and x- projections of the deviance
  computed between the expected and the raw photon counts.
}
  \label{fig:psf:examples}
\end{figure*}


%
We design four digital dictionaries for SBW and WFFM and two dictionaries for
LSCM based on the Gaussian kernel estimated by fitting a SG model parameterised
by $\widehat{\params}^\SG$.
The dictionaries are defined from this estimate by reducing the kernel size, and
placing each kernel at every pixel centre. The parameters for dictionary
$\iDict$ are $\params^{\dict_\iDict} \defeq \set{\widehat{\params}^\SG/\iDict}
\times \set{\x_\iPx}_{\iPx\in\iPxSet}$.
The digital dictionaries allow computing the solution of the LS
sub-problem~\refeq{eq:psf:sp-ls-solution} efficiently with a spectral solver
based on the FFT.
We use reflexive boundary conditions and diagonalise the operators $\vec{O}_k$,
$\transpose{\vec{O}_k}$, and $\transpose{\vec{O}_k}\vec{O}_k$ (thanks to central
symmetry of the Gaussian kernels) using the discrete cosine transform II
\cite{Strang1999,Hansen2006}.
The dictionary parameters are summarised in \reftable{table:psf:dict-designs}.

We vary the amount of regularisation to explore the trade-off between accuracy
and efficiency of the GM model (\reffig{fig:psf:tradeoff:one-kernel}).
We apply \refalg{alg:asb-psf} to estimate the net mixture intensity, obtain the
effective dictionary by thresholding the weights above
$ \wKerMin \defeq y_\MIN c^{-1} \sqrt{8\pi^3 \stdXY^4
  \stdZ^2}$,
where $y_\MIN$ is the $99$th percentile of the Poisson distribution with mean
intensity $c\,\bgPhi$, and debiase the mixture weight using L-BFGS-B.
Examples of the estimated PSF models are shown in \reffig{fig:psf:examples}.
We observe that for $\dict_1$, \ie when the dictionary is built from the kernel
estimated by fitting the SG model to localise the bead, the effective dictionary
is small, and its size barely depends on the regularisation and is comparable
with the SG model in accuracy.
If the dictionary is built from smaller kernels, the regularisation parameter
controls the number of Gaussian kernels: for low regularisation the support of
the mixture weights is large and the model overfits (\eg GM1 for SBW), whereas
increasing the regularisation reduces overfitting while capturing the PSF lobes
(\eg GM2 for SBW).
From the highest density region (HDR) plots \cite{Hyndman1996}, we observe that
the kernels are placed at high-signal regions. When the regularisation
increases, kernels are removed from the lower-signal regions.

These observations are consistent with the literature: in \cite{Zhang2007}, it
has been shown that no accurate Gaussian approximation exists for a 3D WFFM PSF,
whereas an SG approximation is nearly perfect for LSCM.
Comparing the accuracy of the GM and SG models, we observe that the former
brings significant improvement for WFFM and only a slight improvement for LSCM
(\reffig{fig:psf:tradeoff:one-kernel}).

\subsubsection{GM model with a two-kernel dictionary}
\label{sec:psf:tradeoff:model:mg:2k}
We built the two-kernel dictionaries by merging single-kernel dictionaries:
$\dict_2 \cup \dict_3$ for SBW and WFFM, $\dict_1 \cup \dict_2$ for CLSM. We
explore the accuracy--efficiency trade-off by varying the regularisation
parameter of each kernel in the dictionary
(\reffig{fig:psf:tradeoff:two-kernels}).
We notice that similarly to the one-dimensional case, the preference is given to
smaller kernels: they mostly determine the efficiency and the accuracy of the GM
model.
However, adding a larger kernel decreases the effective dictionary size, hence
increasing the efficiency.
We investigate the spatial distribution of each kernel in the dictionary by
computing the empirical probability map that a kernel is selected at a given
pixel among the reconstructions across all regularisation parameter
combinations.
We observe that the larger kernels are selected at a few positions only, near
the centre of the PSF, whereas smaller kernels are mostly positioned at the side
lobes.

\subsubsection{GM model: robustness to the measurement noise}
\label{sec:psf:tradeoff:model:mg:2k}
%

\begin{figure*}[!t]
  \centering
  \includegraphics[width=0.9\textwidth]{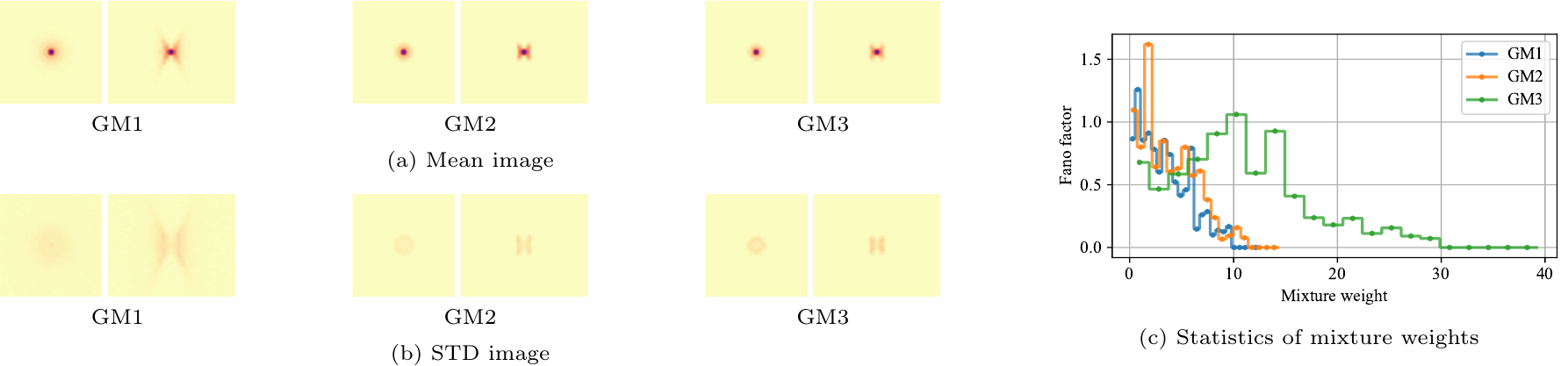}
  \subfloat{\label{fig:psf:robustness-to-noise:mean}}
  \subfloat{\label{fig:psf:robustness-to-noise:std}}
  \subfloat{\label{fig:psf:robustness-to-noise:stats}}
  \caption{
    \textbf{Robustness of the GM PSF model reconstruction to the measurement noise.}
    We estimate the GM PSF models from image stacks of a point source generated using
    the BW PSF model for 50 independent realisations of the noise model.
    We use the same dictionaries and regularisation parameters as for GM1, GM2,
    and GM3 models in \reffig{fig:psf:tradeoff:one-kernel:bw} and
    \reffig{fig:psf:tradeoff:two-kernels:bw}.
    We sample the estimated PSF models and display orthogonal z- and x- mean
    projections of mean \refsubfig{fig:psf:robustness-to-noise:mean} and standard
    deviation \refsubfig{fig:psf:robustness-to-noise:mean} over the image stacks.
    In~\refsubfig{fig:psf:robustness-to-noise:stats}, we show the variability of the
    mixture weights histogram by showing the Fano factor (variance over mean of the
    count in each bin).
    %
    %
  }
  \label{fig:psf:robustness-to-noise}
\end{figure*}


%
We simulate 50 independent realisations of the noise model to generate synthetic
image of an idealised PS object using the BW PSF model as described in
\refsection{sec:psf:imaging-settings:bw}.
We estimate the GM PSF models using the same dictionary and regularisation
parameters as for the GM1, GM2, and GM3 models displayed in
\reffig{fig:psf:tradeoff:one-kernel:bw} and
\reffig{fig:psf:tradeoff:two-kernels:bw}.
As expected, the size of the support and the associated kernel locations in the
estimated PSF models vary across the noise realisations.
To assess the robustness of the GM PSF reconstruction, we investigate the
variability in the estimated mixtures indirectly, by projecting the PSF model on
the pixel grid using \refeq{eq:psf:ifm-gm-vec}, where we set the background
intensity to zero.
We compute the mean and standard deviation over the estimated image stacks
(\reffig{fig:psf:robustness-to-noise:mean} and
\reffig{fig:psf:robustness-to-noise:std}).
We observe that the model consistently captures the overall PSF shape including
the tails (\reffig{fig:psf:robustness-to-noise:mean}).
However, variations remains in the estimated PSF models, mainly in the tails
where the signal is dimmer: in these regions noise has a larger effect on the
position and weight of the estimated Gaussian kernels
(\reffig{fig:psf:robustness-to-noise:std}).

In addition, to assess the robustness in the mixture weights, we compute for
each independent realisation the histogram of the estimated mixture weights
(\reffig{fig:psf:robustness-to-noise:stats}). For each bin we compute the mean
and variance of the number of kernels falling into it and compute the Fano
factor (variance over mean of the count) to capture the variability of the
weight distribution.
We observe that for all models, the kernels with largest weights are estimated
robustly (small Fano factor), whereas the kernels with lower weights display
more variability (large Fano factor). We also note a qualitative difference
between one-dictionary (GM1 and GM2) and two-dictionary (GM3) models. The
two-dictionary model has on average fewer kernels ($24.58 \pm 2.56$), with
higher mixture weights, resulting in a qualitatively different distribution than
the one-dictionary models having on average more kernels (GM1 has $551.68 \pm
16.87$ kernels and GM2 has $98.60 \pm 5.87$ kernels) with lower mixture weights.


\subsection{Application: point source localisation}
\label{sec:psf:ps-loc}
%

\begin{table}[!t]
  \renewcommand{\arraystretch}{1.3}
  \newcommand{\head}[1]{\textnormal{\textbf{#1}}}
  \newcommand{\normal}[1]{\multicolumn{1}{c}{#1}}
  \newcommand{\na}{\emph{n.a.}}

  \colorlet{tableheadcolor}{gray!25} 
  \newcommand{\headcol}{\rowcolor{tableheadcolor}} %
  \colorlet{tablerowcolor}{gray!10} 
  \newcommand{\rowcol}{\rowcolor{tablerowcolor}} %
  \setlength{\aboverulesep}{0pt}
  \setlength{\belowrulesep}{0pt}

  \newcommand*{\rulefiller}{
    \arrayrulecolor{tableheadcolor}
    \specialrule{\heavyrulewidth}{0pt}{-\heavyrulewidth}
    \arrayrulecolor{black}}

  \caption{Accuracy--Efficiency trade-off}
  \label{table:psf:compare}
  \centering
  \begin{threeparttable}
    \begin{tabular}{c*{4}{c}}
      \toprule
      \headcol \head{Model} & \head{Accuracy} & \head{Efficiency} & \multicolumn{2}{c}{\head{Run-time} [\SI{}{\s}]\tnote{a}} \\
      \rulefiller \cmidrule(lr){2-2} \cmidrule(lr){3-3} \cmidrule(lr){4-5}
      \headcol & $\nll$ & $\abs{\supp\paren{\wKer}}$ & mean & std \\
      \midrule
      BW\tnote{b} & $3.55\expe{+5}$ & $-$ & $442.5772$ & $5.226578$  \\
      SG & $3.82\expe{+5}$ & $-$ & $0.031730$ & $0.000878$\\
      SGT & $3.94\expe{+5}$ & $-$ & $0.032802$ & $0.000217$ \\
      GM1 & $3.68\expe{+5}$ & $546$ & $0.155430$ & $0.001401$ \\
      GM2 & $3.78\expe{+5}$ & $93$ & $0.050053$ & $0.001265$ \\
      GM3 & $3.80\expe{+5}$ & $27$ & $0.041315$ & $0.000756$ \\
      \bottomrule
    \end{tabular}
    \begin{tablenotes}
    \item[a] Statistics computed on 100 images of size
      $81 \times 81 \times 101$ pixels. Ran on an Intel(R) i7-4770 CPU (\SI{3}{\GHz}) with 8 cores.
    \item[b] Unoptimised python implementation. For comparison, the
      approximation implemented in \cite{Kirshner2011} yields a typical
      run-time of \SI{2.3}{s} on Intel Core i7 (\SI{2.5}{\GHz}) with 4 cores.
    \end{tablenotes}
  \end{threeparttable}
\end{table}


\begin{figure*}[!t]
  \centering
  \includegraphics[width=0.9\textwidth]{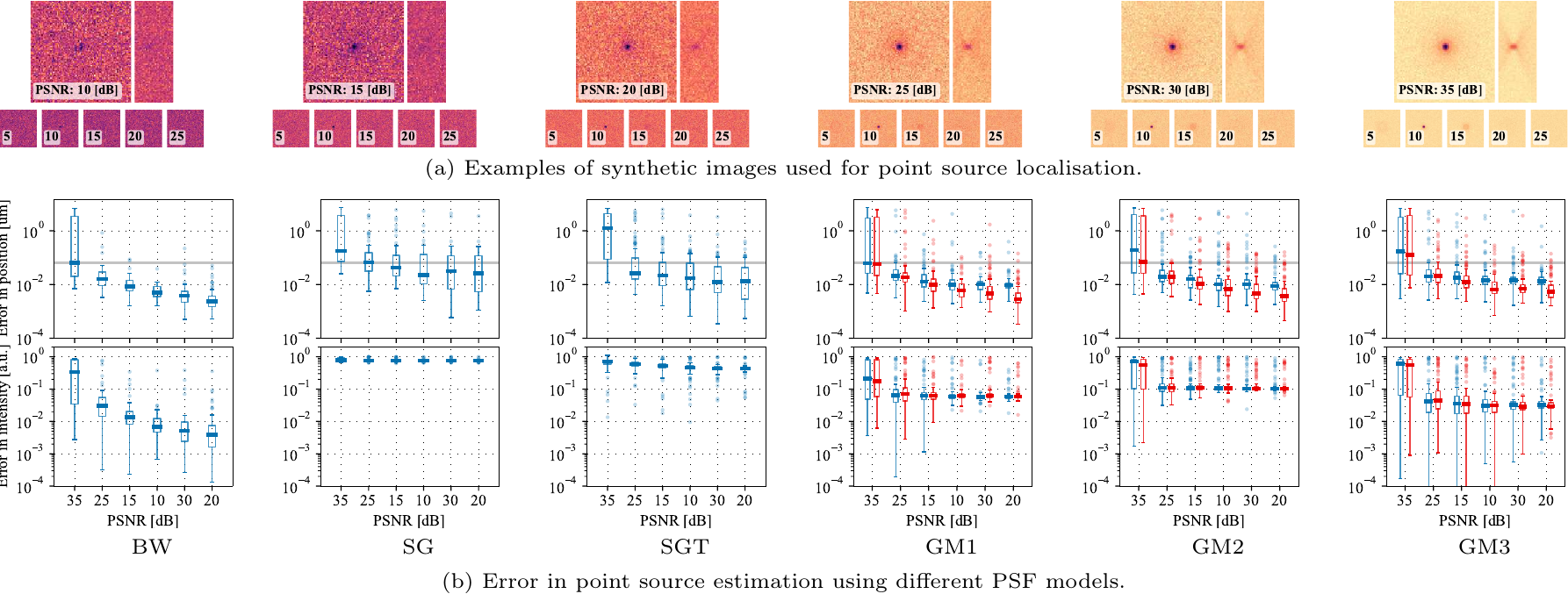}
  \subfloat{\label{fig:psf:compare-models:image-examples}}
  \subfloat{\label{fig:psf:compare-models:results}}
  \caption{
  \textbf{Comparison of the PSF models in solving the PS localisation problem.}
  \refsubfig{fig:psf:compare-models:image-examples} Synthetic image dataset for
  different \psnr of a PS imaged under a BW PSF model. \emph{First row:}
  orthogonal z- and x- projections. \emph{Second row:} image slices acquired at
  different focal planes.
  \refsubfig{fig:psf:compare-models:results} Box plots of the $\ell_2$ error in
  PS position (\emph{first row}) and intensity (\emph{second row}) estimated
  with six PSF models: the ground truth Born and Wolf model (BW); two SG models
  parametrised by either solving the blind PS deconvolution (SG) or derived from
  theory (SGT); three GM models (GM1, GM2, GM3) selected on the SBW dataset (see
  \reffig{fig:psf:examples}).
  The horizontal grey line indicates the pixel size.
  GM models: using $\blurMat_\iKerCov\paren{\psPosHat[]}$ (blue) or
  $\blurMat_\iKerCov\paren{\psPos[\mathrm{gt}]}$ (red) in
  \refsection{sec:psf:op-splitting}.
  %
}
  \label{fig:psf:compare-models}
\end{figure*}


%
We investigate how the PSF model affects the solution to the PS localisation
problem.
To assess quantitatively the performance of the different PSF models, we
generate synthetic image data using the VM framework \cite{Samuylov2015}.
We virtually image a single PS of unit intensity located at a position sampled
uniformly within the imaging volume.
%
%
We vary the exposure time to generate image data at different peak
signal-to-noise ratio (\psnr, adapted from \cite{Aguet2009}):
\begin{equation*}
  \tE = \frac{
  \bgPhi[] + \frac{1}{\nPx} \sum_{\iPx\in\iPxSet}{ \psfFun^\BW_\params \paren{\pxPos - \psPos[]} }
  }{
  \pxArea \max_{ \iPx\in\iPxSet }\paren*{ \psfFun^\BW_\params \paren{\pxPos - \psPos[]} }^2
  } \,
  10^{0.1\,\psnr} \ .
\end{equation*}
The image formation parameters are the same as for the SBW dataset
(\reftable{table:psf:imaging-parameters}).
Sample images are shown in \reffig{fig:psf:compare-models:image-examples}.

We solve the PS localisation problem using CMA-ES for six PSF models: the ground
truth BW; two SG models with parameters estimated either by solving the blind PS
deconvolution problem or derived theoretically (SG, SGT, see
\refsection{sec:psf:tradeoff:model:sg}); three GM models estimated in
\refsection{sec:psf:tradeoff:model:mg:1k} (GM1, GM2) and
\refsection{sec:psf:tradeoff:model:mg:2k} (GM3).
The SG and GM models were calibrated on the SBW data (see
\reffig{fig:psf:psf-gallery:SBW} and \reftable{table:psf:imaging-parameters}).
It is custom to compare PS localisation algorithms only by the position
accuracy. However, we are also interested in the error in intensity.
Therefore, we normalise each PSF models such as to match their central mode,
which is normalised to one.
We summarise the accuracy, the efficiency measures, and the evaluation run-time
for each PSF model in \reftable{table:psf:compare}.

We solve independently the PS localisation problem for $100$ image stacks. We
compute the $\ell_2$ error between the ground truth and the estimated values of
the PS intensity and position (\reffig{fig:psf:compare-models:results}).
We observe that the lowest errors are achieved by the model matching the ground
truth, \ie BW.
However, this accuracy comes at a high computational cost
(\reftable{table:psf:compare}) due to the repeated computations of integrals
(\refsection{eq:psf:model-bw}).
The SG and SGT approximations result in a nanometer-accurate estimate in
position with a low run-time, but at the expense of a higher error in the PS
intensity estimate that does not decrease significantly with higher \psnr.
Extrapolating the results from \refsection{sec:psf:exp-1d-ex}, it is likely that
the SG model captures both the central mode and the side lobes at the expense of
an underestimated central mode amplitude.
For SGT, the central mode is better approximated by design \cite{Zhang2007} and
the error in the estimated intensity decreases, up to a certain extent: for
higher \psnr, the side lobe contribution becomes significant, and the SGT
approximation starts showing its limits.

For each GM model, we solve the PS localisation problem for two cases: when the
mixtures are centred around the estimated position of the fluorescent bead
during calibration $\psPosHat[]$ (see
\refsection{sec:psf:gm:ps-position-estimation} and
\refsection{sec:psf:tradeoff:model:sg}) or centred around the true position
$\psPos[\mathrm{gt}]$, \ie in \refsection{sec:psf:op-splitting} we use
$\blurMat_\iKerCov\paren{\psPosHat[]}$ or
$\blurMat_\iKerCov\paren{\psPos[\mathrm{gt}]}$, respectively.
The former allows quantifying the errors in a real-world scenario, \ie when the
PS position used for calibrating the PSF is unknown. The latter allows
quantifying the errors in an idealised scenario, \ie when the GM model is
optimally calibrated.
In the real-world scenario, the GM models result in position estimates with an
accuracy comparable with the SG and the SGT models. In the idealised scenario,
the accuracy is improved even for GM3, modelled only with 27 Gaussian kernels.
In both scenarios, the three GM models result in a significantly lower error in
the PS intensity estimate compared with SG and SGT.




\section{Discussion}
\label{sec:psf:discussion}
Applying complex PSF models is mainly restricted to digital problems thanks to
the efficiency provided by the FFT \cite{Hansen2006}.
However, for inverse problem requiring an analog reconstruction space, such as
PS localisation or tracking, complex models become infeasible due to their high
evaluation cost.
A common solution is to rely on Gaussian approximations \cite{Zhang2007}.
Nonetheless, several approaches have been proposed to enable complex PSF models
in applications: speeding up the theoretical models by deriving suitable
numerical approximation (\eg approximate complex integrals in the BW model by
truncating a Taylor series, see \cite{Aguet2009}) or by introducing analog
phenomenological approximations with a reasonable computational cost (\eg
B-Splines and Zernike polynomials).
In this work, we propose modelling the PSF using a Gaussian mixture, and we
propose a variational framework to achieve an optimal trade-off between
reconstruction accuracy and computational efficiency.
We believe that we could apply our variational framework to mixtures using other
kernels. However, using a Gaussian kernel has the advantage of subsuming the
customary Gaussian model and to use a computational back-end with an efficient
implementation, \ie the IFGT algorithm \cite{Yang2003}.
However, we made several simplifying assumptions that lay the ground for future
research.

In this paper, we solve the inverse problem for one fluorescent bead, modelled
as a point-like object.
However, reconstructing the PSF given images of several beads scattered in the
image volume can improve the reconstruction accuracy: we expect that such an
extension would improve the sparsity of the GM model by filtering out the false
positive atoms in the dictionary.
It is a straightforward extension that would replace the $\nll$ by a sum of
$\nll$, one for each bead. This extension lends itself to the operator splitting
framework presented in~\refsection{sec:psf:asb-solving}.
However, it requires the beads to be identical and the PSF shift invariant.
In this work, we modelled a bead using an idealised PS. However, the extended
bead size and the asymmetry due to manufacturing imperfections may introduce
artefacts in the PSF reconstruction that affect any subsequent inverse problem.
A non-point-like fluorescent bead can be modelled using the virtual microscope
framework~\cite{Samuylov2015}.
However, the blind PS deconvolution problem will become more involved. We could
potentially explore several extensions: spherical bead with a radius fixed to
the value reported by the manufacturer; spherical bead with a radius estimated
within a range provided by the manufacturer; or non-spherical bead, requiring
shape estimation.
Our framework can be extended to space-varying PSFs. In~\cite{Denis2015},
shift-variant PSFs are modelled as a linear combination of shift-invariant PSFs;
this is already very close to our representation~\refeq{eq:psf:ifm-gm-vec}.
This extension could be accommodated by our framework by extending the
dictionary structure that would account for shift-varying effects.

Our framework applies to both analog and digital dictionaries.
However, for simplicity and efficiency, we used a very specific dictionary
structure, where each Gaussian kernel is placed at every pixel centre. This
allows using a spectral solver based on the FFT for the least-squares problem
(see \refsection{sec:psf:solution:sub-problems} and \refalg{alg:asb-psf}).
The advantage is that the large dictionary matrix is not built and hence not
stored in memory, providing a good memory footprint to the
algorithm~\cite{Hansen2006}.
However, our framework can also handle the more general analog dictionaries,
thanks to IFGT, but at the expense of using a spectral solver for the least
squares problem.
To bypass this limitation, linearising the least-squares problem would result in
an explicit step involving only the evaluation of the GM model, amenable to fast
computation using IFGT, avoiding to build and store $\dict$.
We expect that using such analog dictionaries would allow even sparser
representation of the GM model.
In this paper, we also fix the dictionary structure and kernel parameters.
However, one could formulate a more general inverse problem by optimising for
both the dictionary and the mixture weights.
This would be related to the (non-parametric) dictionary learning
problem~\cite{Mairal2014}, except, we would restrict the dictionary to a
parametric class, \ie Gaussian kernels.
This is in contrast to the usual goal of dictionary learning, that is to the
best of our knowledge, mostly non-parametric.

To explore the accuracy--efficiency trade-off, we used a convex heuristic based
on the $\ell_1$ norm to penalise the number of Gaussian kernels.
This heuristic entails a bias, that results in our case in biased mixture
weights.
To alleviate this problem, we use a straightforward refitting strategy that
requires only solving a maximum likelihood problem, refitting the weights only
for the atoms in the estimated support.
However, the need for debiasing stems from the use of a convex heuristic to
approximate the non-convex cardinality problem.
This non-convex problem could be tackled directly using greedy algorithms such
as orthogonal matching pursuit~\cite{Bach2012}.
Another approach would be to use non-convex approximations, or to explore more
general sparsity-inducing priors.
Both have the advantage of keeping the logic of the operator splitting we
presented here.
Many well-known sparsity-inducing heuristics can be decomposed as a
\emph{difference of convex function}, resulting in an alternating algorithm
solving at each iteration a weighted $\ell_1$ convex
problem~\cite{Gasso2009,Lethi2015}.
This extension would be straightforward, and our algorithm could be reused as a
central component.
These methods feature a better sparsity-inducing behaviour, at the expense of
non-convex algorithms that require a greater care than their convex
counterpart.




\section{Conclusion}

Accurate PSF models are fundamental for solving inverse problems in fluorescence
microscopy.
In practice, Gaussian approximations are favoured for their computational
efficiency, at the expense of their ability to capture the rich structures of
the PSF tails.
In this paper, we introduce a new class of analog PSF models based on a sparse
convex combination of Gaussian mixtures (GM).
We formulate a maximum \aposteriori problem for their calibration from image
data and derive an efficient algorithm based on a \emph{fully-split formulation}
of the \emph{alternating split Bregman} algorithm.
Our formulation allows trading accuracy for efficiency by controlling the number
of Gaussian mixtures modelling the PSF.
We assessed the GM model on synthetic and real image data, and applied it to the
point source localisation problem.
We showed that the GM model allows a good localisation accuracy together with an
improved photometry estimate, at a reasonable computational cost thanks to the
IFGT.
We believe that our framework will contribute to a wider adoption of more
accurate PSF models and hence improve the reconstruction quality of the
intensity signal. This will enable tackling challenging dynamical biological
processes from fluorescence microscopy image data.

\appendix[Notations Summary]

See Table~\ref{table:psf:notations} at the end of the paper.


\begin{table*}[t]

  \caption{Summary of notations used in the paper}
  \label{table:psf:notations}

  \centering

\begin{tabular}[t]{ll}

  \renewcommand{\arraystretch}{1.3}
  \newcommand{\head}[1]{\textnormal{\textbf{#1}}}
  \newcommand{\normal}[1]{\multicolumn{1}{c}{#1}}
  \newcommand{\na}{\emph{n.a.}}

  \colorlet{tableheadcolor}{gray!25} 
  \newcommand{\headcol}{\rowcolor{tableheadcolor}} %
  \colorlet{tablerowcolor}{white}
  \newcommand{\rowcol}{\rowcolor{tablerowcolor}} %
  \setlength{\aboverulesep}{0pt}
  \setlength{\belowrulesep}{0pt}

  \begin{tabular}[t]{lp{5cm}}

    \toprule
    \headcol \head{Notation} & \head{Description} \\
    \midrule\rowcol
    $\x \defeq (x,y,z) \in \R^3$ & Arbitrary point in physical space \\
    $\psPos[] \in \R^3$ & Position of a point source \\
    $\psPhi[] \in \Rpos$ & Photon emission rate of a point source \\
    $\bgPhi[] \in \Rpos$ & Photon emission rate of the background \\
    $\dx$ & Lebesgue measure in space \\
    $\dt$ & Lebesgue measure in time \\
    $\dirac_{\x}\paren{\dx}$ & Dirac measure positioned at $\x$ \\
    $\objMeas[]^\OBJECT\paren{\dy \times \dt}$ & Spatio-temporal object measure modelling the total photon flux emitted by a point source object and the background \\
    $\tE$ & Exposure time \\
    $\imageDom \subset \R^3$ & Imaging volume \\
    $\pxArea$ & Pixel area \\
    $\lateralSampling$ & Pixel size \\
    $\nHeight \times \nWidth$ & Camera detector size (in pixels) \\
    $\axialSampling$ & Distance between two focal planes \\
    $\nSlices$ & Number of focal planes \\
    $\obsGVMat[] \in \Zpos^{\nSlices\times\nHeight\times\nWidth}$ & Image stack data (grey values) \\
    $\iPxSet \defeq \set{1, \dots, \nPx}$ & Linear indices over $\obsGVMat[]$, with $\nPx \defeq \nSlices\nHeight\nWidth$\\
    $\px_\iPx \subset \imageDom$ & Surface of the $\iPx$-th pixel in physical space\\
    $\x_{\iPx} \in \imageDom$ & Centre of the $\iPx$-th pixel in physical space\\
    $\expPhCount[\iPx]\paren{\objMeas[]}$ & Expected photon count at the $\iPx$-th pixel \\
    $\obsPhCount[\iPx]$ & (Random) photon count at the $\iPx$-th pixel \\
    $\obsGVCount[\iPx]$ & (Random) grey value at the $\iPx$-th pixel \\
    $\mapPxToImg$ see eq.~\eqref{eq:psf:px2img} & Pixel-to-image mapping modelling the conversion of photons hitting the detector surface $\obsPhCount[\iPx]$ into grey values $\obsGVCount[\iPx]$ \\
    $\lambda$ & emission wave length \\
    $\quantEfficiency$ & quantum efficiency at $\wavelength$ \\
    $\mGain$ & multiplication gain\\
    $\ADU$ & analog-to-digital proportionality factor \\
    $\cameraBias$ & camera bias \\
    $\NA$ & Numerical aperture \\
    $\refIndImm$ & Refractive index of the immersion medium \\
    $\psfFun\paren{\dx \mid \y}$ & Transition probability kernel \\
    $\psfFun_{\params}\paren{\y}$ & PSF kernel with parameters $\params \in \paramSet$ \\
    $\params^\BW \defeq \set{\wavelength, \NA, \refIndImm}$ & Parameters of the Born and Wolf model \\
    $\params^\SG \defeq \set{\stdXY, \stdZ}$ & Lateral and axial standard deviations parametrising single Gaussian PSF models \\
    $\Sigma_{\params} \in \R^{3\times3}$ & Covariance matrix of a Gaussian kernel \\
    $\dict$ & Parametric dictionary describing Gaussian mixture (GM) PSF model, where each atom is a shifted SG kernel\\
    $\abs{\dict}$ & Dictionary size, \ie the number of Gaussian kernels in the
                    mixture \\
    $\params^\dict$ & Parameters of the kernels in $\dict$ \\
    $\iKerCovSet \defeq \set{1, \dots, \nKerCov}$ & Index set of the covariance matrices in $\dict$ \\
    $\kerPosSet_\iKerCov$ with $\iKerCov \in \iKerCovSet$
    & Set of $\nKerPos$ positions in $\dict$ associated with a Gaussian kernel parameterised by $\params^\SG_\iKerCov$ \\
    $\iKerPosSet_\iKerCov \defeq \set{1, \dots, \nKerPos}$ & Indices over $\kerPosSet_\iKerCov$ \\
    \bottomrule
  \end{tabular}

  \begin{tabular}[t]{lp{5cm}}
    \toprule
    \headcol \head{Notation} & \head{Description} \\
    \midrule\rowcol
    $\simplex_{\abs{\dict}}$ &  Probability simplex defined as the convex hull of the $\abs{\dict}+1$ standard unit vectors\\
    $\wKerNormedVec[] \in \simplex_{\abs{\dict}}$ & Vector of mixture weights $\wKerNormed$ with $\iKerCov \in \iKerCovSet$ and $\iKerPos \in \iKerPosSet_\iKerCov$  \\
    $\wKerNormedVec \in \Rpos^{\nKerPos}$ & Vector obtain by grouping the weights associated with a Gaussian mixture parameterised by $\params_\iKerCov$: $\parenEl{\wKerNormedVec}_\iKerPos \defeq \wKerNormed$\\
    $\wKerNetVec[\iKerCov] \in \Rpos^{\nKerPos}$ & Net mixture intensity defined as $\psPhi[] \wKerNormedVec$ \\
    $\wKerNetVec \in \Rpos^{\abs{\dict}}$ & Stacked vector of net mixture intensities \\
    $\supp(\wKerNormedVec[])$ & Support of the mixture weights defined as a set of indices of non-zero mixture weights \\
    $\params^\GM \defeq \params^\dict \cup \set{\wKerNormedVec[]}$ & Parameters of the Gaussian mixture model \\
    $\nll$ & Data fitting term defined as a normalised negative log-likelihood, also called in statistics deviance or Bregman divergence, see~\cite{Paul2013} \\
    $\nllNormed$ & $\nll$ scaled by $\nPx$ \\
    $\regTerm$ & Regularisation term (prior about the model parameters)\\
    $\reg$ & Regularisation parameter controlling the trade-off between $\nll$ and $\regTerm$ \\
    $\regVec \in \Rpos^\nKerCov$ & Vector of $\nKerCov$ regularisation parameters associated with a set of kernels parameterised by $\params^\SG_\iKerCov$ \\
    $\obsPhCountVec[] \in \Rpos^{\nPx}$ & Vectorised raw photon counts obtained by applying the inverse of $\mapPxToImg$ to $\obsGVMat[]$ \\
    $\expPhCountVec[] \in \Rpos^{\nPx}$ & Vector of expected photon counts computed according to object-to-pixel mapping at each pixel $\iPx \in \iPxSet$ \\
    $\psfVec[\iKerCov]\paren{\x} \in \Rpos^{\nKerPos}$ & Vector obtained by evaluating a Gaussian mixture parameterised by $\params_\iKerCov$ at $\x - \x_{\iKerCov\iKerPos}$ with $\x_{\iKerCov\iKerPos} \in \kerPosSet_\iKerCov $ \\
    $\blurMat_\iKerCov\paren{\x} \in \Rpos^{\nPx \times \nKerPos}$ & Blurring matrix where each row is defined as $\psfVec[\iKerCov]\paren{\x_{\iPx} - \x}$ at each pixel $\iPx \in \iPxSet$ \\
    $\zerosVec_{n} \in \R^{n}$ & $n$-dimensional vector of zeros \\
    $\onesVec_{n} \in \R^{n}$ & $n$-dimensional vector of ones \\
    $\eyeMat_{n} \in \R^{n \times n}$ & Identity matrix \\
    $\indicatorOptim_\mathcal{S}\paren{\wKer}$ & Indicator functional that assumes $0$ if $\wKer$ is componentwise non-negative and $+\infty$ otherwise \\
    \bottomrule
  \end{tabular}

\end{tabular}
\end{table*}


\section*{Acknowledgment}
\addcontentsline{toc}{section}{Acknowledgment}

We thank T. Julou, X. Chen and Y. Barral for sharing the orginal bead
measurements. This work has been supported by the SystemsX.ch RTD
Grant \#2012/192 TubeX of the Swiss National Science Foundation.

\bibliographystyle{IEEEtran}
\bibliography{IEEEabrv,bib-psf/refs,bib-psf/books}

\begin{IEEEbiography}[{\includegraphics[width=1in,height=1.25in,clip,keepaspectratio]{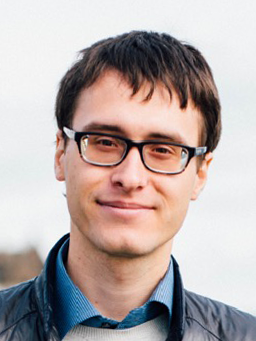}}]{Denis
    K. Samuylov}
  received the B.Sc. degree in telecommunications from the Saint Petersburg
  State Polytechnic University (currently the Peter the Great Saint Petersburg
  Polytechnic University), Russia, in 2011. He received the M.Sc. degree in
  interdisciplinary approaches to life science from the Paris Diderot University
  (Paris 7), France, in 2013. He received the Ph.D. degree in computer vision
  from the Computer Vision laboratory at the Swiss Federal Institute of
  Technology in Zurich (ETH Zurich), Switzerland, in 2018. His research
  interests include the application of signal processing and computer vision
  techniques to various problems in biology and medicine.
\end{IEEEbiography}

\begin{IEEEbiography}[{\includegraphics[width=1in,height=1.25in,clip,keepaspectratio]{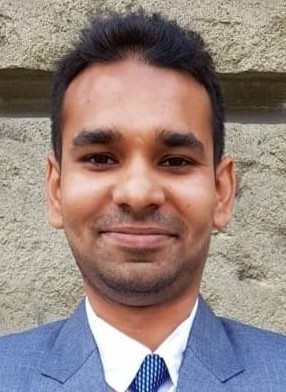}}]{Prateek Purwar}
  Prateek Purwar received the B.Tech. degree in Electrical Engineering from Indian Institute of Technology Bombay, Mumbai, India in 2009. He received the M.Sc. degree in Electrical Engineering from Swiss Federal Institute of Technology in Zurich (ETH Zurich), Switzerland, in 2017. His research interests mainly include the application of signal processing, machine learning and computer vision techniques for solving various problems in medicine and automation industry.
\end{IEEEbiography}

\begin{IEEEbiography}[{\includegraphics[width=1in,height=1.25in,clip,keepaspectratio]{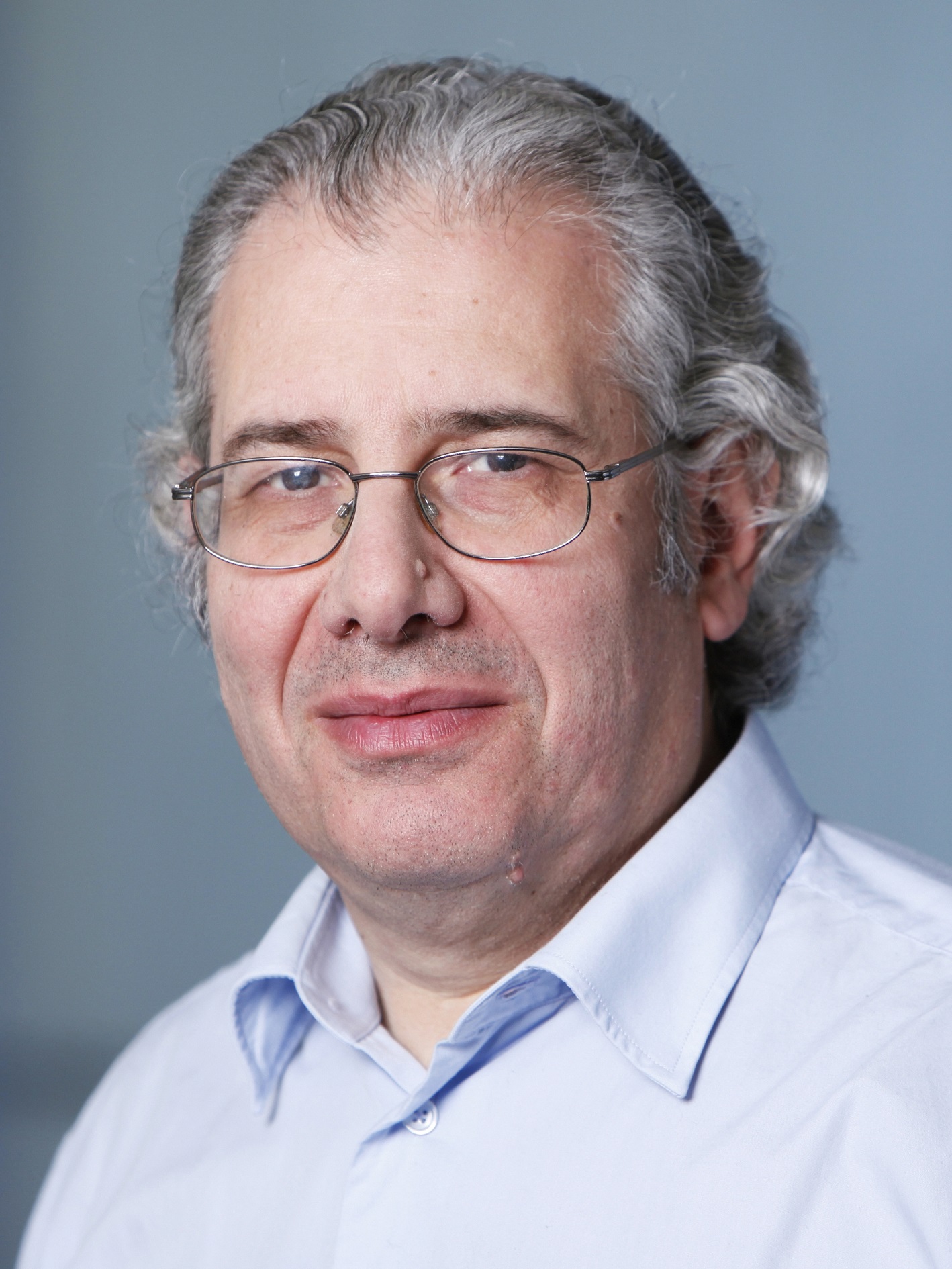}}]{Gábor
    Székely}
  received the Graduate degree in chemical engineering, the Graduate degree in
  applied mathematics, and the Ph.D. degree in analytical chemistry from the
  Technical University of Budapest and the Eötvös Lórand University, Budapest,
  Hungary, in 1974, 1981, and 1985, respectively. Since 2002 he has been leading
  the Medical Image Analysis and Visualization Group at the Computer Vision
  Laboratory of Swiss Federal Institute of Technology (ETH) Zurich, Switzerland,
  concentrating on the development of image analysis, visualization, and
  simulation methods for computer support of biomedical research, clinical
  diagnosis, therapy, training and education.
\end{IEEEbiography}

\begin{IEEEbiography}[{\includegraphics[width=1in,height=1.25in,clip,keepaspectratio]{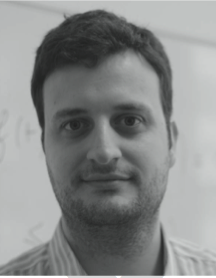}}]{Grégory
    Paul}
  received the M.Sc. degree in cell biology and physiology from Ecole Normale
  Supérieure, Paris, France, in 2003, and the Ph.D. degree from the University
  of Paris VI, in 2008. In 2008 he joined the Swiss Federal Institute of
  Technology (ETH) Zurich, Switzerland, as a post-doctoral researcher in the
  computer science department with Prof. Ivo F. Sbalzarini to develop new
  quantitative tools to investigate biological processes from image data.
  Together with Prof. Gábor Székely, between 2012 and 2017 he was leading the
  BioimagE Analysis and Modeling (BEAM) team, at ETH in the Computer Vision
  Laboratory, focusing on the development of image analysis, computational
  statistics and biophysical modeling applied to cell biology.
\end{IEEEbiography}

\vfill

\end{document}